\documentclass[amsfonts, amssymb, amsmath, reprint, showkeys, nofootinbib, onecolumn]{revtex4-2}
\usepackage[english]{babel}
\usepackage[utf8]{inputenc}
\usepackage{comment}

\usepackage{amsthm}
\usepackage{mathtools}
\usepackage{xcolor}
\usepackage{color}
\usepackage{graphicx}
\usepackage{adjustbox}
\usepackage{placeins}
\usepackage[T1]{fontenc}
\usepackage{lipsum}
\usepackage{csquotes}
\usepackage[dvipdfmx]{hyperref}
\usepackage{bm}
\usepackage{mathrsfs}
\usepackage{amsfonts}
\usepackage{braket}
\begin{document}
\title{
Geometric Steady-State Thermodynamics Engine under an Isothermal Operation 
}

\author{Ryosuke Yoshii}\email[e-mail address:]{ryoshii@rs.socu.ac.jp}
\affiliation{Center for Liberal Arts and Sciences, Sanyo-Onoda City University,
Yamaguchi 756-0884, Japan,}
\affiliation{International Institute for Sustainability with Knotted Chiral Meta Matter (WPI-SKCM2), Hiroshima University, Higashi-Hiroshima, Hiroshima 739-8526, Japan}

\author{Hisao Hayakawa}\email[e-mail address:]{hisao@yukawa.kyoto-u.ac.jp}
\affiliation{Center of Gravitational Physics and Quantum Information, Yukawa Institute for Theoretical Physics, Kyoto University, Kitashirakawa-oiwake cho, Sakyo-ku, Kyoto 606-8502, Japan}

\date{\today}

\begin{abstract}

Geometric pumping in open quantum systems is commonly described in terms of
the Berry--Sinitsyn--Nemenman (BSN) curvature, which determines the
geometric contribution to transported quantities under cyclic parameter
modulation.
In this work, we show that a cyclically driven open quantum system can
operate as an isothermal engine through a mechanism that is
fundamentally distinct from curvature-driven pumping.
In the adiabatic limit, the instantaneous pumping current vanishes,
yet a finite work per cycle survives.
We demonstrate that this work originates from the parametric dependence
of the instantaneous steady state rather than from the BSN curvature.

Using a general superoperator formulation, we derive the non-adiabatic
expansion of the density matrix and separate reversible and
irreversible contributions to work, heat, and entropy production.
The entropy production per cycle scales linearly with the
operation speed, ensuring reversibility in the strict adiabatic limit.
Finite-speed corrections are expressed in terms of a thermodynamic
metric defined on the steady-state manifold, leading to a geometric
bound on efficiency degradation.

As a concrete example, we apply the theory to the Anderson impurity
model under cyclic modulation of electrochemical potentials and 
the strength of Coulomb interaction. 
In the sequential-tunneling regime, we obtain finite work in the non-adiabatic regime and confirm that finite work persists in the adiabatic limit.

These results clarify the geometric structure underlying
isothermal cyclic thermodynamics and reveal a class of 
reversible steady-state engines whose operation is controlled
by the geometry of the steady-state manifold rather than 
by the BSN curvature.

\end{abstract}

\maketitle

\section{INTRODUCTION}\label{sec:intro}

Geometric effects, mainly, in driven open quantum systems have attracted sustained attention in the context of Thouless pumping or adiabatic pumping~\cite{thouless1,thouless2,berry,xiao,sinitsyn1,sinitsyn2}.
When external parameters are varied cyclically and slowly, 
a finite transported quantity per cycle can emerge even in the absence
of a bias. 
This phenomenon is commonly described in terms of the Berry--Sinitsyn--Nemenman (BSN) connection and its associated curvature, which determine the geometric contribution to pumping currents. 
This phenomenon has been experimentally validated in processes such as charge transport \cite{ex-ch1,ex-ch2,ex-ch2.5,ex-ch3,ex-ch4,ex-ch5,ex-thou1,ex-thou2} and spin pumping~\cite{ex-spin1}. 
Theoretical studies have explored this effect using diverse methodologies, including scattering theories \cite{brouwer,s-th1,s-th2,s-th3,s-th-ch1,s-th-ch2,s-th-ch3,s-th-spin1}, classical master equations \cite{parrondo,usmani,astumian1,astumian2,rahav,chernyak1,chernyak2,ren,sagawa,ville2021}, and quantum master equations \cite{qme1,qme2,qme-spin1,qme-spin2,yuge1,yuge2}.
Extended fluctuation theorems for geometric pumping have also been investigated \cite{watanabe,Hino-Hayakawa,Takahashi20JSP}, enriching the theoretical framework.
In such systems, the instantaneous current vanishes in the strict adiabatic limit, while the integrated pumped quantity remains finite
because it is expressed as a geometric surface integral in parameter space. 

The geometric framework has naturally been extended to thermodynamic settings, leading to proposals of geometric heat engines and adiabatic quantum pumps operating under cyclic modulation of reservoir parameters and Hamiltonian couplings~\cite{Wang2021,Brandner-Saito,engine,Hino2021,Bhandari20,Abiuso20,Alonso21,Eglinton,Yoshii22,Yoshii13,Wang2024,Breuer2016,FHHT,Funo2020}.
In most of these constructions, the engine performance is understood as a consequence of the BSN curvature.
The geometric phase accumulated during the cycle is regarded as the fundamental origin of work extraction. 

In this work, we show that this viewpoint is incomplete.
We demonstrate that a cyclically driven open quantum system can operate
as a reversible isothermal engine even when the BSN pumping current
vanishes in the adiabatic limit. 
The extracted work does not originate from the curvature of the BSN connection, but from the parametric dependence of the instantaneous
steady state itself.

The key observation is the following.
In the slow modulation regime, the density matrix can be expanded as
\begin{equation}\label{rho_exp}
\hat{\rho}(\theta)
=
\hat{\rho}^{\mathrm{ss}}(\theta)
+
\epsilon \hat{\rho}^{(1)}(\theta)
+\epsilon^2 \hat{\rho}^{(2)}(\theta)
+O(\epsilon^3),
\end{equation}
where $\hat{\rho}^{\mathrm{ss}}(\theta)$ with the phase $\theta$ characterizing the cyclic modulation is the instantaneous steady state. 
Here, we have introduced the dimensionless operation speed, $\epsilon:= 1 / (\tau_{\mathrm{p}} \Gamma)$, where the parameter modulation with the period $\tau_\mathrm{p}$ commences at $t = 0$, and $\Gamma$ represents the coupling strength or the characteristic transition rate between the system and its reservoirs. 
The BSN curvature governs the first-order correction
$\hat{\rho}^{(1)}$, which generates pumping currents proportional
to the operation speed.
However, the work $W$ performed over one cycle is already determined by the zeroth-order steady-state contribution through
\begin{equation}
W :=\oint
\mathrm{Tr}
\left[
\hat{\rho}(\theta)
\, \frac{\partial}{\partial\theta} \hat{H}(\theta)
\right]
=
\oint
\mathrm{Tr}
\left[
\hat{\rho}^{\mathrm{ss}}(\theta)
\, \frac{\partial}{\partial \theta} \hat{H}(\theta)
\right]
+ O(\epsilon) ,
\end{equation}
where we have introduced $\oint\cdot:= \int_\mathrm{\theta_\mathrm{in}}^{2\pi+\theta_\mathrm{in}}d\theta \cdot$ as a cyclic integral over the period starting from the initial phase $\theta_\mathrm{ini}$.
Because $\hat{\rho}^{\mathrm{ss}}(\theta)$ depends on the electrochemical potentials and parameters to control Hamiltonian $\hat{H}(\theta)$, its cyclic modulation produces finite work even when the geometric pumping current disappears.
The engine, therefore, operates through a reversible modulation of the steady-state manifold rather than through a geometric phase associated with the BSN curvature.

This distinction has important thermodynamic consequences.
In the strictly adiabatic limit, the deviation from the steady state vanishes and the entropy production or the product of the Kullback–Leibler (KL) divergence~\cite{KL51,Sagawa20} per cycle scales as $O(\epsilon)$, as will be shown.
The engine, thus, becomes reversible as $\epsilon \to 0$. 
The efficiency approaches a maximum value and is determined solely by the geometry of the cyclic path in parameter space.
Finite-speed corrections introduce entropy production that can be expressed in terms of a thermodynamic metric on the steady-state manifold, yielding a geometric bound on efficiency degradation.

We also argue for the fast-modulation case, similar to Floquet theory.
As will be shown in Sec. \ref{sec:fast_modulation}, the density matrix can be expressed as
\begin{align}\label{eq:fast}
    \hat{\rho}(\theta)=\hat{\rho}(\theta_\mathrm{ini})+\epsilon^{-1}\hat{\rho}^{(-1)}(\theta)+O(\epsilon^{-2}), 
\end{align}
where $\hat{\rho}^{(-1)}(\theta)$ will be determined later.
Remarkably, the density matrix is unchanged from the initial condition, i.e., $\hat{\rho}(\theta)=\hat{\rho}(\theta_\mathrm{ini})$ in the fast modulation limit.
This helps us to make the problem simple, as will be seen in Sec.~\ref{sec:fast_modulation}.

To make these statements concrete, we formulate a general superoperator framework for slowly and fast driven open quantum systems and derive the adiabatic expansion of the density matrix.
We clarify the thermodynamic structure of work, heat, and entropy production in the isothermal setting and separate reversible and irreversible contributions explicitly.
We then apply the theory to the Anderson impurity model under cyclic modulation of electrochemical potentials and level energies~\cite{Yoshii22,Yoshii13}, in the quasi-classical regime. 

The central message of this work is therefore conceptual: the geometric structure relevant for cyclic thermodynamics is not the BSN curvature itself, but the geometry of the steady-state manifold.
The present engine exemplifies a class of reversible isothermal machines whose operation persists in the strict adiabatic limit, revealing a thermodynamic mechanism distinct from conventional geometric pumping.

The organization of this paper is as follows. 
In Sec.~\ref{sec:general}, we explain the setup and a geometric formulation for describing the heat engine under a slow modulation process, including the descriptions of the first law of thermodynamics in our system. 
In Sec.~\ref{sec:fast_modulation}, we formulate geometric thermodynamics in a fast modulation process.
In Sec.~\ref{sec:appli}, we apply our formulation to the Anderson model for a quantum dot coupled to two reservoirs within the wide-band approximation.
In Sec.~\ref{sec:results}, we present numerical results for the Anderson model.
In Sec.~\ref{discussion}, 
we discuss the underlying physical mechanism of our geometric isothermal engine, contrasting it with conventional curvature-driven pumping, and explore its extensions to many-body correlation regimes and nonlinear transport. 
Finally, in Sec.~\ref{sec:conclusion}, we summarize our results and perspectives with some discussions. 
In Appendix \ref{app:slow-driving}, we briefly summarize the mathematical properties of our general framework.
In Appendix \ref{app:proof_W_minus_one}, we prove that the work becomes positive in the fast modulation case.
In Appendix \ref{app:eigen_modes}, we explicitly write the eigenvalues and eigenvectors of the Anderson model.
In Appendix \ref{app:high_T}, we develop the high-temperature expansion for our modulating systems.
In Appendix \ref{app:proofs_efficiency}, we present the proofs for the signs of the work and the finite speed correction of the efficiency.

\section{GENERAL FRAMEWORK}\label{sec:general}

In this section, we develop a general framework for describing the thermodynamics of a quantum system under adiabatic pumping. We first outline the system setup (Section \ref{sec:setup}) and then derive the first law of thermodynamics in the context of geometric cyclic states (Section \ref{susec:FirstLaw}).

\subsection{Setup}\label{sec:setup}


\begin{figure}
\centering
\includegraphics[clip,width=8cm]{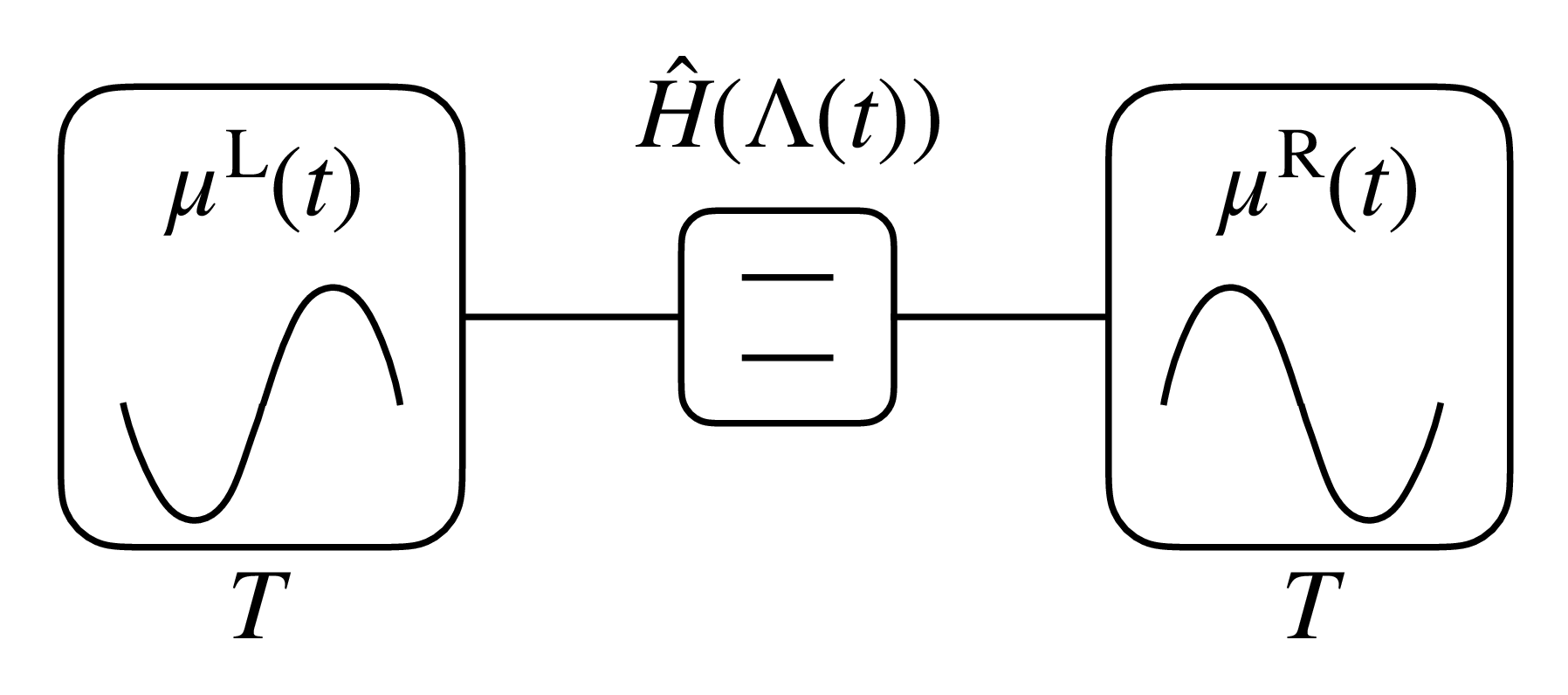}
\caption{
A schematic of the model we consider in this paper, in which $\mu^{\rm L}(t)$, $\mu^{\rm R}(t)$, and $\hat{H}(\Lambda(t))$ are modulated by an external agent. }
\label{fig0}
\end{figure}


In this paper, we consider a small quantum system S such as a quantum dot coupled to two reservoirs L and R under periodic modulation of some system parameters, with the period $\tau_{\mathrm{p}}$. 
We focus on the isothermal condition throughout this study. 
Each reservoir $\alpha=$L or R is characterized by the electrochemical potential $\mu^{\alpha}$ and temperature $T$ (or the inverse temperature $\beta:=1/T$).
In this paper, we adopt the unit with the Boltzmann constant $k_B = 1$ and the Planck constant $\hbar=1$.

Although our general formalism can be applied to many systems, we specify the model employed in this paper for the reader. 
We illustrate the setup by a schematic picture in Fig.~\ref{fig0}, where we modulate $\mu^{\rm L}(t)$ and $\mu^{\rm R}(t)$ in the reservoirs in time $t$ with different phases. 
In this paper, we control the set of parameters 
\begin{equation}\label{control_parameters}
\bm{\Lambda} (t):= (\Lambda_1(t),\Lambda_2(t),\Lambda_3(t)):=
\left(
\Lambda (t),
\frac{\mu^{\mathrm{L}}(t)}{\overline{\mu^{\rm L}}},
\frac{\mu^{\mathrm{R}}(t)}{\overline{\mu^{\rm R}}} 
\right) ,
\end{equation}
where we have introduced
\begin{equation}
\overline{\mu^\alpha}:=\frac{1}{\tau_p}\int_0^{\tau_p}dt\mu^\alpha(t)
\end{equation}
for $\alpha=$L and R. 
Here, $\Lambda(t)$ denotes a parameter inside the quantum dot.\footnote{Though we consider a single parameter inside the quantum dot depends on time here, it is straightforward to generalize to the case where multiple parameters inside the quantum dot depend on time. }
When we consider the case without any difference in time-averaged electrochemical potentials, i.e.\, $\overline{\mu^\mathrm{R}}=\overline{\mu^\mathrm{L}}$, 
 we can replace $\overline{\mu^\alpha}$ in Eq.~\eqref{control_parameters} by $\overline{\mu}$, which is independent of $\alpha$. 
 For later discussion We sometimes write $\bm{\Lambda}(t):=(\Lambda_\nu(t))$ with $\nu=1,2,3$.

We assume that the quantum master equation describes the dynamics of S
\begin{align}\label{master}
    \frac{\partial}{\partial\theta} |\hat{\rho}(\theta)\rangle 
    = \epsilon^{-1} \hat{K}(\bm{\Lambda}(\theta)) |\hat{\rho}(\theta)\rangle,
\end{align}
where $\hat{\rho}(\theta)$ is the density matrix of the system S. 
\footnote{We adopt a simplified notation $\bm{\Lambda}(\theta)=\bm{\Lambda}(t)$.}
In this study, we introduce the dimensionless time, defined as the phase of the modulation, $\theta:= 2\pi t / \tau_{\mathrm{p}}$, and the dimensionless operation speed, $\epsilon:= 1 / (\tau_{\mathrm{p}} \Gamma)$, where the parameter modulation commences at $t = 0$, and $\Gamma$ represents the coupling strength or the characteristic transition rate between the system and its reservoirs. 
For simplicity, we drop the arguments of the density matrix except for $\theta$, though it depends on the various parameters as $|\hat \rho(\theta, \bar  \mu, \cdots )\rangle $.
A previous work~\cite{Yoshii22} investigated relaxation from a nonequilibrium steady state to a geometric state, demonstrating that such relaxation generates entropy production and enables work extraction. 
Additionally, we note that the dynamics governed by physically relevant quantum master equations are completely positive and trace-preserving (CPTP), as established in prior studies~\cite{Lindblad76, Manzano20, Sagawa20}. 
Furthermore, we assume that the master equation, Eq.~\eqref{master}, is of the Markovian type, although the modulation incorporated in $\hat{K}(\bm{\Lambda}(\theta))$ through $\bm{\Lambda}(\theta)$ may introduce non-Markovian effects, as discussed in Ref.~\cite{Mizuta21}. 
\footnote{
Gorini–Kossakowski–Sudarshan–Lindblad equation with non-negative jump rates is CP-divisible even if some jump rates are negative.
However, some convolution-less master equations are non-Markovian, and time-dependent jump rates might be negative~\cite{Breuer2016}.
}
Given our focus on dynamics under slow modulations, we posit that the dynamic evolution of the density matrix $\hat{\rho}(\theta)$ is primarily driven by variations in $\bm{\Lambda}(\theta)$. 
When the diagonal elements of the density matrix $\hat{\rho}(\theta)$ are decoupled from the off-diagonal elements, the system exhibits classical behavior; however, if the diagonal and off-diagonal elements are coupled, the system is inherently quantum.

In Eq.~\eqref{master} we have introduced the supervector consisting of the elements of the density matrix
\begin{equation}\label{rho_vec}
|\hat{\rho}(\theta)\rangle := 
\begin{pmatrix}
\rho_{11}(\theta)\\
\rho_{12}(\theta)\\
\dots,\\
\rho_{NN}(\theta)\\
\end{pmatrix}
, 
\end{equation}
where $\rho_{ij}(\theta)$ is the $(ij)-$element of $\hat{\rho}$ at $\theta$, which is assumed to be a $N\times N$ matrix. 
We note that the diagonal element appears at the $M+N(M-1)$-th component ($1\le M\le N$). 
The density matrix  $\hat{\rho}(\theta)$ satisfies the condition of  probability conservation ${\rm Tr}\hat{\rho}(\theta)=1$.
Thus, the transition matrix $\hat{K}(\bm{\Lambda}(\theta))$ is a superoperator acting on $|\hat{\rho}(\theta)\rangle$, which is expressed as a $N^2\times N^2$ matrix.

We assume that the master equation (\ref{master}) has a unique steady state $|\hat{\rho}^{\mathrm{ss}}(\bm{\Lambda}(\theta)) \rangle$ which satisfies 
\begin{equation}\label{steady_condition}
\hat{K}(\bm{\Lambda}(\theta))|\hat{\rho}^{\mathrm{ss}}(\bm{\Lambda}(\theta)) \rangle = 0.
\end{equation}
Equation \eqref{steady_condition} means that the steady density matrix is equivalent to the right zero eigenstate of $\hat{K}(\bm{\Lambda}(\theta))$.  
We also write
\begin{align}\label{eigen_eq}
    \hat{K}(\bm{\Lambda}(\theta))|r_n(\bm{\Lambda}(\theta)) \rangle=\lambda_n | r_n(\bm{\Lambda}(\theta)) \rangle
\end{align}
as the eigenvalue equation, where $\lambda_n$ and $| r_n(\bm{\Lambda}(\theta)) \rangle$ are $n$--th eigenvalue and right eigenvector, respectively.
Note that the eigenvalues of $\hat K(\bm{\Lambda}(\theta))$ do not explicitly depend on $\epsilon$.
Since the system is coupled to two reservoirs having different electrochemical potentials,  $|\hat{\rho}^{\mathrm{ss}}(\bm{\Lambda}(\theta)) \rangle$ is a nonequilibrium steady state in the presence of the bias voltages. 
In the following, we denote the left zero eigenvector of $\hat{K}(\bm{\Lambda}(\theta))$ as $\langle \ell_0|$, which satisfies $\langle \ell_0|\hat{\rho}^{\mathrm{ss}}(\bm{\Lambda}(\theta)) \rangle={\rm Tr}\hat {\rho}^{\mathrm{ss}}(\bm{\Lambda}(\theta)) =1$. 
This transition matrix has $n^2$ eigenvalues $\lambda_i$ and corresponding eigenvectors. 
Since the transition matrix is non-Hermitian, the left eigenstates $\langle \ell_i|$ and right eigenstates $|r_i\rangle$ with the eigenvalue $\lambda_i$ are not the Hermitian conjugate. 
For later convenience, we label the eigenvalues $\lambda_i$ such that $\lambda_0=0>\mathrm{Re}[\lambda_1] \ge\mathrm{Re}[\lambda_2]\ge \cdots \ge\mathrm{Re}[\lambda_{n^2-1}]$, where $\mathrm{Re}[\lambda_i]$ represents the real part of $\lambda_i$. 
For simplicity, we assume that the degeneracy exists only in the $\mathscr{M}$-th excited state with well-separated eigenvalues as $\lambda_{\mathscr{M}}-\lambda_{\mathscr{M}-1}\gg \epsilon$ with $\mathscr{M}\le N^2-1$.
In this convention, the 0-th eigenvector coincides with the steady state vector  $|r_0\rangle=|\hat{\rho}^{\mathrm{ss}}(\bm{\Lambda}(\theta)) \rangle$.

\subsection{First law of thermodynamics}
\label{susec:FirstLaw}

Now, we discuss the thermodynamic relations for our system. 
The energy of the subsystem is given by 
\begin{equation}
E(\theta):=\mathrm{Tr}[\hat H(\Lambda(\theta))\hat\rho(\theta)]. 
\end{equation}
Its rate of change satisfies the first law
\begin{align}
\frac{\partial E(\theta)}{\partial \theta}
=
\mathscr{P}(\theta)+\mathscr{J}(\theta),
\label{dE_re}
\end{align}
where the power $\mathscr{P}(\theta)$ and heat current $\mathscr{J}(\theta)$ are, respectively, defined as
\begin{align}
\mathscr{P}(\theta)
&:=
{\rm Tr}\left[
\hat{\rho}(\theta)\frac{\partial \hat{H}(\Lambda(\theta))}{\partial \theta}
\right],
\qquad 
\mathscr{J}(\theta)
:=
{\rm Tr}\left[
\hat{H}(\Lambda(\theta)) \frac{\partial\hat{\rho}(\theta)}{\partial\theta}
\right] .
\label{J(theta)}
\end{align}

For $\epsilon\ll 1$, the density matrix is expanded as Eq.~\eqref{rho_exp}. Thus, the explicit forms of $\mathscr{P}^{(0)}(\theta)$ and $\mathscr{P}^{(1)}(\theta)$ in $\mathscr{P}(\theta)=\mathscr{P}^{(0)}(\theta)+\epsilon \mathscr{P}^{(1)}(\theta)+O(\epsilon^2)$ are expressed as
\begin{align}\label{P_0,P_1}
    \mathscr{P}^{(0)}(\theta)={\rm Tr}\left[
\hat{\rho}^\mathrm{ss}(\theta)\frac{\partial \hat{H}(\Lambda(\theta))}{\partial \theta}\right], \quad 
\mathscr{P}^{(1)}(\theta)={\rm Tr}\left[
\hat{\rho}^{(1)}(\theta)\frac{\partial \hat{H}(\Lambda(\theta))}{\partial \theta}
\right].
\end{align}
Also, using  Eq.~\eqref{pn} with $n=1,2$ (see Appendix \ref{app:slow-driving} for the derivation), we can write
\begin{align}
\hat{\rho}^{(1)}(\theta)
=
\hat{K}^+(\bm{\Lambda}(\theta))
\frac{\partial \hat{\rho}^{\rm ss}}{\partial\theta},
\qquad
\hat{\rho}^{(2)}(\theta)
=
\left[\hat{K}^+(\bm{\Lambda}(\theta))
\frac{\partial}{\partial\theta}
\right]^2 \hat{\rho}^{\rm ss}(\theta), 
\label{rho1_def_re}
\end{align}
where we have introduced
\begin{align}\label{K+}
    \hat{K}^{+}(\bm{\Lambda}) := \sum_{m\neq 0} \varepsilon_{m}(\bm{\Lambda})^{-1} |r_{m}(\bm{\Lambda})\rangle \langle \ell_{m}(\bm{\Lambda})| .
\end{align}

Substituting Eq.~\eqref{rho1_def_re} into the heat current Eq.~\eqref{J(theta)}, we obtain
\begin{align}
\mathscr{J}(\theta)
&=\mathscr{J}^{(0)}(\theta)+\epsilon \mathscr{J}^{(1)}(\theta) +O(\epsilon^2)\notag\\
&= {\rm Tr}\left[
\hat{H}\frac{\partial \hat{\rho}^{\rm ss}}{\partial \theta}
\right]+\epsilon{\rm Tr}\left[
\hat{H}
\frac{\partial }{\partial \theta}\left(\hat{K}^{+}\frac{\partial \hat{\rho}^{\rm ss}}{\partial \theta} \right)
\right]+O(\epsilon^2).
\label{J_geometric}
\end{align}
Remarkably, the heat current $\mathscr{J}^{(n)}(\theta)$ is determined by $(\hat{K}^+\partial_\theta)^n\hat{\rho}^{\rm ss}(\theta)$.

The heat over a cycle is defined as
\begin{align}\label{def:Q}
Q
:=
\oint  \, \mathscr{J}(\theta).
\end{align}

Similarly, the work is given by
\begin{align}\label{def:W}
W
:=
\oint  \, \mathscr{P}(\theta)
=
\oint \,
\mathrm{Tr}\left[
\hat{\rho}(\theta)
\frac{\partial \hat{H}}{\partial \theta}
\right].
\end{align}

Defining the absorbed and released heat as
\begin{align}\label{Q_A&Q_R}
Q_{\rm A}:=\oint \mathscr{J}\Theta(\mathscr{J}),
\quad
Q_{\rm R}:=-\oint \mathscr{J}\Theta(-\mathscr{J}),
\end{align}
where we have introduced Heavisine's step function $\Theta(\mathscr{J})=1$ if and only if $\mathscr{J}\ge 0$, the heat can be expressed as 
\begin{align}
Q=Q_\mathrm{A}-Q_\mathrm{R} .    
\end{align}
Accordingly, we define efficiency:
\begin{align}\label{def:eta}
    \eta:=-\frac{W}{Q_\mathrm{A}} .
\end{align}

\subsection{Geometrical entropy production and thermodynamic length}

We now formulate the entropy production in a way consistent with the above dynamical description.

We introduce the KL divergence $D^{\mathrm{KL}}(\hat\rho||\hat{\rho}^{\rm ss})$ as
\begin{align}\label{def:KL}
D^{\mathrm{KL}}(\hat\rho||\hat{\rho}^{\rm ss})
:=
\mathrm{Tr}\left[\hat\rho (\ln \hat\rho -\ln \hat{\rho}^{\rm ss})\right].
\end{align}
The non-adiabatic entropy production rate is defined as
\begin{align}
\dot\sigma
:=
- \frac{\partial}{\partial\theta}
D^{\mathrm{KL}}(\hat\rho||\hat{\rho}^{\rm ss}).
\end{align}
Substituting the expansion Eq.~\eqref{rho_exp}, we obtain
\begin{align}
\dot{\sigma}^{(1)}={\rm Tr}
\left[
\hat{K}(\bm{\Lambda})
\hat{\rho}^{(1)}
(\hat{\rho}^{\rm ss})^{-1}
\hat{\rho}^{(1)}
\right], 
\label{sigma_main}
\end{align}
where we have used $\sigma=\epsilon \sigma^{(1)}+O(\epsilon^2)$, i.e., $\Sigma=\epsilon \Sigma^{(1)}+O(\epsilon^2)$.
Namely, there is no entropy production in the adiabatic limit:
\begin{equation}\label{Sigma_ad=0}
\Sigma \to \Sigma^{(0)}= 0, \qquad \sigma^{(0)}=0 .
\end{equation}
Thus, the entropy production is expressed solely in terms of 
$\hat{\rho}^{(1)}$, the same quantity that determines the heat current.

Using Eq.~\eqref{rho1_def_re}, we write
\begin{align}
\hat{\rho}^{(1)}
=
\partial_\mu \hat{\rho}^{\rm ss}
\, \dot{\Lambda}_\mu,
\end{align}
where
\begin{align}
\partial_\mu \hat{\rho}^{\rm ss}
:=
\hat{K}^+ \frac{\partial}{\partial \Lambda_\mu}
\hat{\rho}^{\rm ss}.
\end{align}

Then, the entropy production $\Sigma\approx \epsilon \Sigma^{(1)}$ over one cycle becomes
\begin{align}
\Sigma^{(1)}
&=
\oint \, \dot{\sigma}^{(1)}(\theta)
=
\oint \,
|\lambda_1(\theta)|
g_{\mu\nu}(\bm{\Lambda}(\theta))
\dot{\Lambda}_\mu \dot{\Lambda}_\nu ,
\end{align}
where the metric tensor is given by
\begin{align}
g_{\mu\nu}(\bm{\Lambda})
:=
\frac{1}{2}
{\rm Tr}
\left[
\partial_\mu \hat{\rho}^{\rm ss}
(\hat{\rho}^{\rm ss})^{-1}
\partial_\nu \hat{\rho}^{\rm ss}
\right].
\end{align}
This metric defines the thermodynamic length
\begin{align}
\mathcal{L}
:=
\oint
\sqrt{
|\lambda_1|
g_{\mu\nu}
d\Lambda_\mu d\Lambda_\nu
}.
\end{align}

Using the Cauchy--Schwarz inequality, i.e., $\int |f|^2d\theta\int |g|^2d\theta\ge (\int |fg|d\theta)^2$ with $f=\sqrt{ |\lambda_1|g_{\mu\nu}\dot \Lambda_\mu\dot \Lambda_\mu}$ and $g=1$, we obtain
\begin{align}
2\pi\Sigma^{(1)}
\ge
 \mathcal{L}^2\ge 0.
\label{CS_re}
\end{align}
This is a remarkable result because the KL divergence defined as Eq.~\eqref{def:KL} is non-monotonic as demonstrated in Ref.~\cite{Yoshii22}, i.e. $\dot\sigma$ may change the sign at finite $\epsilon$, while $\Sigma^{(1)}$ becomes non-negative.

\subsection{Efficiency: Adiabatic and Non-Adiabatic Regimes}

\subsubsection{Adiabatic regime}

In the limit $\epsilon \to 0$, the system remains on the instantaneous steady state,
\begin{equation}
\hat{\rho}(\theta) = \hat{\rho}^{\rm ss}(\theta) + O(\epsilon),
\end{equation}
and the entropy production vanishes as shown in Eq.~\eqref{Sigma_ad=0}.

Thus, the process is reversible. Over one cycle, the internal energy is periodic, and we obtain
\begin{align}
\oint \frac{\partial E}{\partial \theta}
=0,
\end{align}
which implies
\begin{align}\label{Q+W}
Q^{(0)} + W^{(0)} = 0, 
\end{align}
where we have used expansions $Q=Q^{(0)}+\epsilon Q^{(1)}+O(\epsilon^2)$ and $W=W^{(0)}+\epsilon W^{(1)}+O(\epsilon^2)$.

Here, we write $Q^{(n)}$ and $W^{(n)}$ with $n=0$ and 1 explicitly as
\begin{align}\label{Q_0,Q_1}
    Q^{(0)}&=\oint {\rm Tr}\left[
\hat{H}(\Lambda(\theta)) \frac{\partial\hat{\rho}^\mathrm{ss}(\theta)}{\partial\theta} 
\right] 
, \quad
Q^{(1)}=\oint {\rm Tr}\left[
\hat{H}(\Lambda(\theta)) \frac{\partial\hat{\rho}^{(1)}(\theta)}{\partial\theta} 
\right] =
\oint {\rm Tr}\left[
\hat{H}(\Lambda(\theta)) \frac{\partial}{\partial\theta}\hat{K}^+\partial_\theta\hat{\rho}^\mathrm{ss}(\theta) 
\right]
, \\
W^{(0)}&=\oint \,
\mathrm{Tr}\left[
\hat{\rho}^\mathrm{ss}(\theta)
\frac{\partial \hat{H}}{\partial \theta}
\right] ,
\quad
W^{(1)}=\oint \,
\mathrm{Tr}\left[
\hat{\rho}^{(1)}(\theta)
\frac{\partial \hat{H}}{\partial \theta}
\right] ,
\label{W_0,W_1}
\end{align}
where we have used Eqs.~\eqref{J(theta)}, \eqref{rho1_def_re}, \eqref{def:Q}, and \eqref{def:W}.

Thus, the adiabatic efficiency $\eta_\mathrm{ad}$ satisfies
\begin{align}\label{Carnot}
\eta_\mathrm{ad}
:=
-\frac{W^{(0)}}{Q_{\rm A}^{(0)}}
=
1-\frac{Q_{\rm R}^{(0)}}{Q_{\rm A}^{(0)}}
\le 1 ,
\end{align}
where we have used the expansions $Q_\mathrm{R}=Q_\mathrm{R}^{(0)}+\epsilon Q_\mathrm{R}^{(1)}+O(\epsilon^2)$ and $Q_\mathrm{A}=Q_\mathrm{A}^{(0)}+\epsilon Q_\mathrm{A}^{(1)}+O(\epsilon^2)$.
This relation is analogous to the expression for the Carnot efficiency.
We can evaluate $W^{(0)}$ and $Q^{(0)}$ easily, which will be explained later separately, while it is hard to distinguish $Q_\mathrm{A}^{(0)}$ from $Q_\mathrm{R}^{(0)}$ without observation.
In this paper, we adopt the following strategy: (i) we calculate $\mathscr{J}(\theta)$ theoretically, and (ii) we use Eq.~\eqref{Q_A&Q_R} to evaluate $Q_\mathrm{A}$ and $Q_\mathrm{R}$.  

\subsubsection{Non-adiabatic regime}

For nonzero $\epsilon$, the non-adiabatic density matrix $\hat{\rho}^{(1)}$ generates entropy.
The total heat can be decomposed into reversible and irreversible parts:
\begin{align}
Q^{(1)}
=
Q^{(1)}_{\rm rev}
+
Q^{(1)}_{\rm irr},
\end{align}
where the irreversible part is related to entropy production as
\begin{align}
Q^{(1)}_{\rm irr}
=
-T \Sigma^{(1)} \le 0. 
\end{align}
This means that $Q^{(1)}_{\rm irr}$ is a part of $Q_\mathrm{R}^{(1)}$.

Thus, over one cycle, we obtain the relation 
\begin{align}\label{Eq40}
Q^{(1)}_\mathrm{rev}+W^{(1)}=T\Sigma^{(1)} .
\end{align}
Using this, the non-adiabatic efficiency defined as
\begin{align}\label{eta_nonad}
\eta_\mathrm{nad}&:=\eta-\eta_\mathrm{ad}
\end{align}
is reduced to
\begin{align}\label{eta^{(1)}}
\eta^{(1)}&= - \left\{\frac{W^{(1)}}{Q_\mathrm{A}^{(0)}}-\frac{Q_\mathrm{A}^{(1)}W^{(0)}}{Q_\mathrm{A}^{(0) 2}} \right\} 
\le \frac{T\Sigma^{(1)}-W^{(1)}-T\mathcal{L}^2/(2\pi)}{Q_A^{(0)}}+\frac{Q_\mathrm{A}^{(1)}W^{(0)}}{Q_\mathrm{A}^{(0) 2}} ,
\end{align}
where $\eta^{(1)}:=\lim_{\epsilon\to 0}\eta_\mathrm{nad}/\epsilon$.
Note that both Eqs. \eqref{Carnot} and \eqref{eta_nonad} contain $Q_\mathrm{A}^{(i)}$ and $Q_\mathrm{R}^{(i)}$ with $i=0, 1$, although we know $W^{(i)}$ and $\mathcal{L}$.
When we evaluate them, we check whether $Q$ is absorbing or releasing during the time evolution.
If we are allowed to observe the sign of $Q$, we can determine the efficiency theoretically.
We expect $\eta^{(1)}$ should be $\eta^{(1)}<0$.
This is supported by $Q_\mathrm{A}^{(i)}>0$, $W^{(0)}<0$ in a thermodynamic engine and $\mathcal{L}^2>0$ if $Q_\mathrm{rev}^{(1)}$ is not extremely large.

\section{Fast modulation limit}\label{sec:fast_modulation}

In the fast modulation limit $\epsilon^{-1}\to 0$, we cannot use Eq.~\eqref{rho_exp} for a slow modulation process. 
In this case, we may ignore the change of the density matrix during the operation.
Thus, the density matrix is unchanged from the initial state.
This can be verified from
\begin{align}
    \lim_{\epsilon^{-1}\to 0}\frac{\partial}{\partial\theta} |\hat{\rho}(\theta)\rangle=\lim_{\epsilon^{-1}\to 0} \epsilon^{-1}\hat{K}(\bm{\Lambda}(\theta)) |\hat{\rho}(\theta)\rangle=0. 
\end{align}
If we take an initial state as a nonequilibrium steady state $|\hat{\rho}^\mathrm{ss}(\theta_\mathrm{ini})\rangle$, the density matrix can be approximated as 
\begin{align}\label{fast_linear}
    |\hat \rho(\theta)\rangle= |\hat \rho^{\rm ss}(\theta_\mathrm{ini})\rangle+\frac{1}{\epsilon}\int_{\theta_\mathrm{ini}}^\theta\hat{K}(\bm{\Lambda}(\phi))d\phi |\hat \rho^{\rm ss}(\theta_\mathrm{ini})\rangle+O(\epsilon^{-2}) . 
\end{align}
Thus, if we ignore the contribution of $O(\epsilon^{-1})$, we can ignore the time evolution of the density matrix as explained in Eq.~\eqref{eq:fast}.

\subsection{Work under fast modulation}

In the fast modulation limit, the work generated in a single cycle becomes zero from Eq.~\eqref{def:W} since  
\begin{align}\label{W_fastlimit}
    \lim_{\epsilon^{-1}\to0}W={\rm Tr}\left[\hat\rho^{\rm ss}(\theta_\mathrm{ini})\oint \frac{\partial \hat H(\theta)}{\partial \theta } \right]=0.  
\end{align}
Since $\hat{\rho}(\theta)$ is unchanged in the fast modulation limit, $\lim_{\epsilon^{-1}\to 0}\mathscr{J}=0$ and $\lim_{\epsilon^{-1}\to 0}Q=0$.
Thus, the efficiency is ill-defined in the fast modulation limit.

\subsection{The finite speed correction based on $\epsilon^{-1}$ expansion}

As mentioned in the previous subsection, the time evolution in the vicinity of the initial state is approximated by Eq.~\eqref{fast_linear} for $\epsilon\gg 1$, although the correction of $O(\epsilon^{-1})$ is based on a heuristic argument. 
Here, we develop the formal $\epsilon^{-1}$ expansion to account for the finite-speed effect. 

Assuming the insensitivity of $\hat{\rho}(\theta)$ during the time evolution, Eq.~\eqref{master} can be rewritten as
\begin{align}
    |\hat \rho(\theta)\rangle&\simeq |\hat \rho(\theta-\Delta \theta)\rangle+\frac{1}{\epsilon}\int^\theta_{\theta-\Delta \theta}\hat{K}(\bm{\Lambda}(\phi))d\phi |\hat \rho(\theta-\Delta \theta)\rangle+O(\epsilon^{-2},\Delta\theta^2)\nonumber\\ 
    &=S(\theta;\theta-\Delta \theta)|\hat \rho(\theta-\Delta \theta)\rangle+O(\epsilon^{-2},\Delta\theta^2) ,
    \label{epsiloninvexpansion2}
\end{align}
where we have introduced 
\begin{align}
    \label{epsiloninvexpansion3}
    &\hat{S}(a;b):=1+\frac{1}{\epsilon}\int^a_b\hat{K}(\bm{\Lambda}(\phi))d\phi. 
\end{align}
The recursive expression of Eq.~\eqref{epsiloninvexpansion3} is given as 
\begin{align}
    &|\hat \rho(\theta)\rangle\simeq \hat S(\theta;\theta-\Delta \theta)\hat S(\theta-\Delta\theta;\theta-2\Delta \theta)\cdots \hat S(\Delta \theta;0)|\hat \rho(0)\rangle+O(\epsilon^{-2},\Delta\theta^2).
    \label{epsiloninvexpansion4} 
\end{align}

Equation~\eqref{epsiloninvexpansion4} can be rewritten as 
\begin{align}
    &|\hat \rho(\theta)\rangle\simeq e^{\frac{1}{\epsilon}\int_{\theta_\mathrm{ini}}^\theta\hat{K}(\bm{\Lambda}(\phi))d\phi}|\hat \rho(\theta_\mathrm{ini})\rangle+O(\epsilon^{-2}) ,
    \label{epsiloninvexpansion5} 
\end{align}
where we have used
\begin{align}
    \lim_{\Delta\theta\to 0}S(a ;a -\Delta\theta)&=\lim_{\Delta\theta\to 0}\left[1+\frac{1}{\epsilon}\int^a_{a-\Delta \theta}\hat{K}(\bm{\Lambda}(\phi))d\phi\right]\nonumber\\
    &= e^{\frac{1}{\epsilon}\int^a_{a-\Delta \theta}\hat{K}(\bm{\Lambda}(\phi))d\phi}. 
\end{align}
Equation \eqref{epsiloninvexpansion5} reduces to $|\hat{\rho}(\theta)\rangle=|\hat{\rho}(\theta_\mathrm{ini})\rangle$ in the limit $\epsilon^{-1} \to 0$.
Using $\epsilon^{-1}$ expansion, we obtain
\begin{align}\label{rho^{-}}
    |\hat{\rho}^{(-1)}(\theta)\rangle=\int_{\theta_\mathrm{ini}}^{\theta}\hat{K}(\bm{\Lambda}(\phi))d\phi |\hat \rho^{\rm ss}(\theta_\mathrm{ini})\rangle ,
\end{align}
as mentioned in Eq. \eqref{fast_linear}.

Substituting Eqs.~\eqref{eq:fast} and \eqref{rho^{-}} into Eq.~\eqref{def:W}, we obtain $W=\epsilon^{-1}W^{(-1)}+O(\epsilon^{-2})$ as
\begin{align}\label{W-}
W^{(-1)}=\oint \langle\ell_0|\dot {\hat H}\left\{\int_{\theta_\mathrm{ini}}^{\theta}\hat{K}(\bm{\Lambda}(\phi))d\phi \right\} |\hat \rho^{\rm ss}(\theta_\mathrm{ini})\rangle . 
\end{align}
Note that the positivity of $W^{(-1)}$ can be proven as shown in Appendix \ref{app:proof_W_minus_one}.
Similarly, one can calculate the heat current  $\mathscr{J}=\epsilon^{-1}\mathscr{J}^{(-1)}+O(\epsilon^{-2})$ and the heat generated in a single cycle $Q=\epsilon^{-1}Q^{(-1)}$ as 
\begin{align}\label{Q-}
\mathscr{J}^{(-1)}= \langle\ell_0|\hat H\hat{K}(\bm{\Lambda}(\theta)) |\hat \rho^{\rm ss}(\theta_\mathrm{ini})\rangle,\qquad  Q^{(-1)}=\oint  \mathscr{J}^{(-1)} d\theta .
\end{align}
Note that Eqs.~\eqref{W-} and \eqref{Q-} generate non-zero $W^{(-1)}$ and $Q^{(-1)}$ because $K(\bm{\Lambda}(\theta))$ acts on $\hat{\rho}^\mathrm{ss}(\theta_\mathrm{ini})$.

We observe that the non-commutativity of $\hat{K}$ at different times is negligible in the fast modulation limit, whereas it becomes significant for finite $\epsilon$. 
Given the analytical difficulty of evaluating $W$, $Q$, and $\eta$ precisely at $O(\epsilon^{-1})$, we restrict our analysis to demonstrating that $W$ scales proportionally with $\epsilon^{-1}$ in the fast modulation regime.
It should be noted that $W$ is positive for large $\epsilon$, and thus, we cannot extract the work as an engine if we modulate parameters fast.

\section{APPLICATION to a quantum dot system}\label{sec:appli}

\begin{figure}
\centering
\includegraphics[clip,width=8cm]{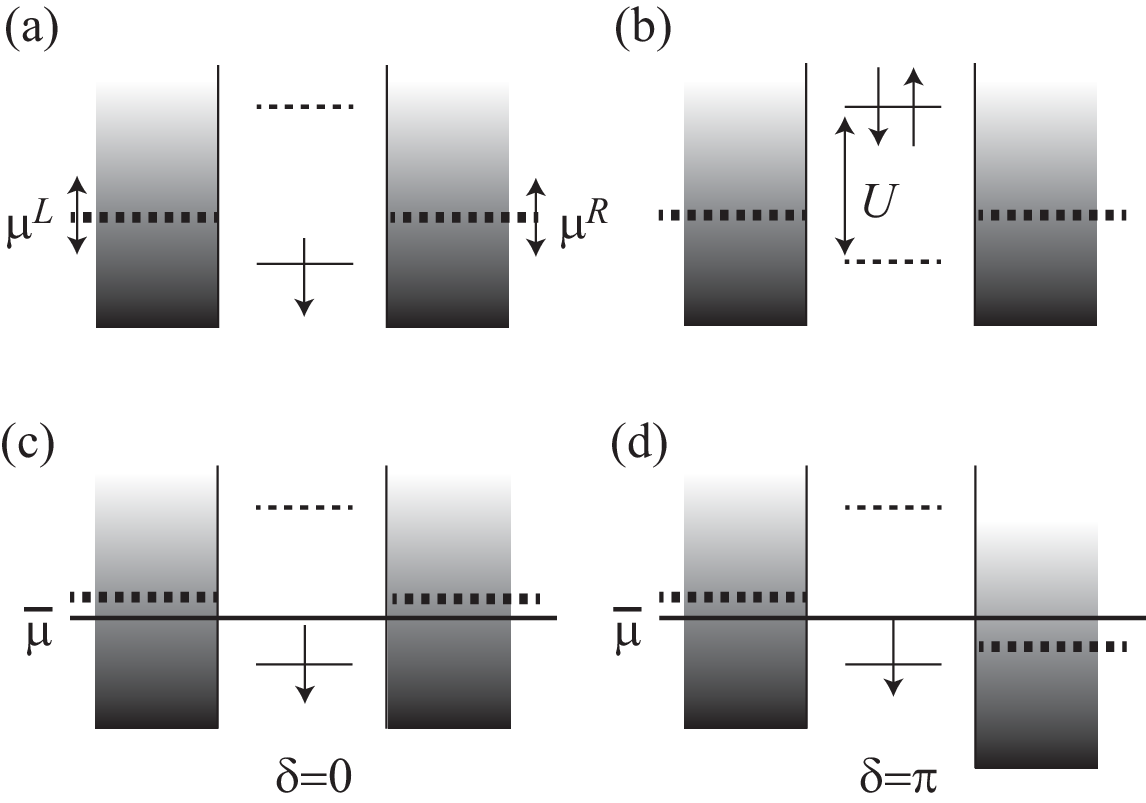}
\caption{
(a) 
Schematics of a modulation of the Anderson model in which two reservoirs are connected to a quantum dot.
We modulate the electrochemical potentials in the left ($\mu^{\rm L}(\theta)$) and right  ($\mu^{\rm R}(\theta)$) reservoirs. 
(b)
We also modulate the Coulomb repulsion $U$ inside the quantum dot as a tunable parameter.
Another control parameter, $\delta$, represents the phase difference between the modulations of the two electrochemical potentials. 
Figures (c) and (d) are schematics for $\delta=0$ and $\delta=\pi$, respectively.    
}
\label{figAM}
\end{figure}

In this section, we apply the general framework developed in the previous sections to the Anderson model for a quantum dot (QD) in which a single dot is coupled to two electron reservoirs.
Figure \ref{figAM} illustrates the time evolution of our system.


The total system consists of the single-dot system and baths (reservoirs).
Thus, the total Hamiltonian $\hat{H}^{\rm tot}$ is written as 
\begin{align}\label{H_tot}
\hat{H}^{\rm tot}:&=\hat{H}+\hat{H}^{\rm r}+\hat{H}^{\rm int} ,
\end{align}
where the system Hamiltonian $\hat{H}$, reservoir Hamiltonian $\hat{H}^{\rm r}$ and interaction Hamiltonian $\hat{H}^{\rm int}$ are, respectively, given by
\begin{align}\label{H_s}
\hat{H}&=\sum_\sigma \epsilon_0 \hat{d}_\sigma^\dagger \hat{d}_\sigma+U(\theta) \hat{n}_\uparrow \hat{n}_\downarrow .
\\
\label{H_bath}
\hat{H}^{\rm r}&=\sum_{\alpha,k,\sigma}\epsilon_k \hat{a}_{\alpha,k,\sigma}^\dagger \hat{a}_{\alpha,k,\sigma}
\\
\hat{H}^{\rm int}&=\sum_{\alpha,k,\sigma}V_\alpha \hat{d}_\sigma^\dagger \hat{a}_{\alpha,k,\sigma}+{\rm h.c.},
\label{H_int}
\end{align}
where $\hat{a}_{\alpha, k,\sigma}^\dagger$ and $\hat{a}_{\alpha, k,\sigma}$ are, respectively, the creation and annihilation operators for the electron in the reservoirs $\alpha$$=$(L or R) with wave number $k$, energy $\epsilon_k$, and spin $\sigma=(\uparrow$ or $\downarrow$).
Moreover,  $\hat{d}^\dagger_\sigma$ and $\hat{d}_\sigma$ are those in the QD, and  
$\hat{n}_{\sigma}=\hat{d}^\dagger_{\sigma}\hat{d}_{\sigma}$. 
Here, we assume that $U(\theta)$ is controlled as
\begin{equation}\label{U(theta)}
U(\theta)=U_0\Lambda(\theta).
\end{equation}
We have introduced the transfer energy $V_\alpha$ between the QD and the reservoir $\alpha$ in Eq.~\eqref{H_int}. 
We adopt a model in the wide-band limit for the reservoirs.
In this paper, the line width is given by $\Gamma=\pi \mathfrak{n} V^2$, where $V:=\sqrt{V_L^2+V_R^2}$ and $\mathfrak{n}$ is the density of states in the reservoirs. 
For simplicity, we focus on the case of $V_L=V_R$ throughout this study.

As shown in Appendix~\ref{app:eigen_modes} with explicit expressions for eigenvalues and eigenvectors, this system is non-degenerate, and off-diagonal elements of the density matrix do not play any role.
In particular, the explicit expression of $\hat{\rho}^\mathrm{ss}(\theta)=r_0(\theta)$, which plays an important role in the analysis, is presented in Eq.~\eqref{right_zero}. 
Since the system is essentially classical, the transition matrix $\hat{K}(\bm{\Lambda}(\theta))$ satisfies the properties of transition matrices for classical stochastic processes.

\section{Results in the quasi classical regime}\label{sec:results}

\subsection{Setup}

In this section, we calculate the thermodynamic quantities discussed in Sec.~\ref{sec:general} numerically with the aid of the detailed properties of the Anderson model.  
In this paper, we consider geometric work generation caused by a modulation of the parameters. 
$\overline{\mu}:=\overline{\mu^{\mathrm{\alpha}}}$ and $U(\theta)$, and fix the other parameters.
To remove the effect of the initial relaxation process, we begin with a steady state $\hat{\rho}^\mathrm{ss}$ at $\theta=0$ and observe the time evolution after one cycle, i.e., we set $\theta_\mathrm{ini}=2\pi$ throughout our analysis.

Although Ref.~\cite{Abiuso20} discussed the optimal path for high efficiency and work, we adopt the following control of the set of parameters 
$\bm{\Lambda}(\theta)$ as
\begin{align}\label{3rd_protocol}
    \Lambda(\theta) &= 1 + r_{\lambda} \cos\theta, \\
    \frac{\mu^{\mathrm{L}}(\theta)}{\overline{\mu}} &= 1 + r_\mu \sin \theta, \\
    \frac{\mu^{\mathrm{R}}(\theta)}{\overline{\mu}} &=1+ r_\mu \sin(\theta+ \delta), 
\end{align}
where $\delta$  is the phase difference between the electrochemical potentials in the left and right reservoirs.

If we take $\delta \neq 0$, the electrochemical potential difference between the two reservoirs remains nonzero. 
In this case, the dynamic current proportional to chemical potential differences (bias voltage) is much larger than the geometric current.
Although the average of the dynamic current over one cycle is zero, the average of the geometric current remains nonzero.
If we consider $\delta=0$, there exists only the geometric current.
Thus, the geometric effect plays an important role in our system.

\begin{figure}
\centering
\includegraphics[clip,width=18.5cm]{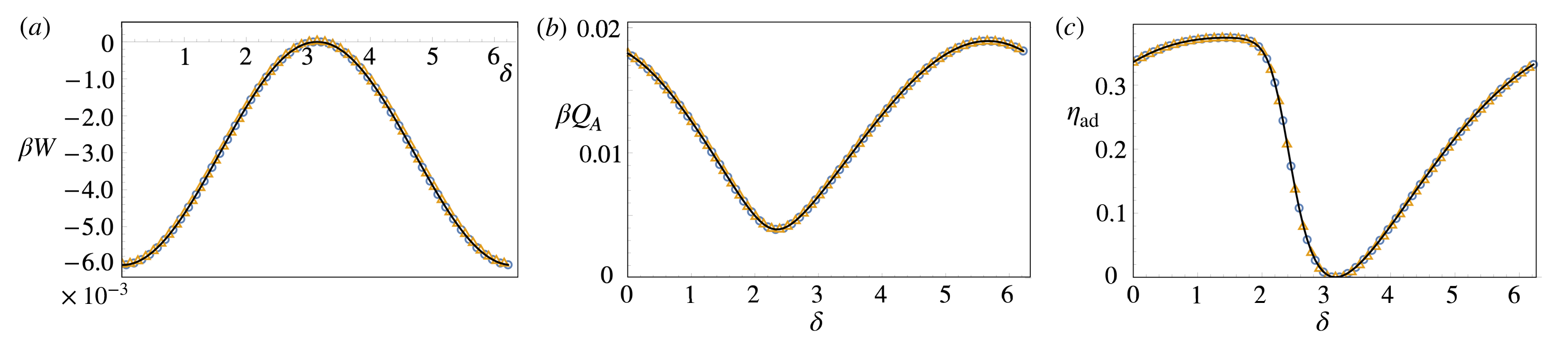}
\caption{Plots of $\beta W$ (a) and $\beta Q_{\rm A}$ (b) and $\eta$ (c) as functions of $\delta$ for $r=0.9$, $\beta U_0=0.1$, $\beta \epsilon_0=0.1$, $\theta_\mathrm{ini}=2\pi$, and $\beta \bar \mu =0.1$. 
The solid lines represent the theoretical results for $W^{(0)}$, $Q_{\rm A}^{(0)}$ and $\eta_{\rm ad}$ in the adiabatic limit (see Eqs. \eqref{Q_0,Q_1}, \eqref{W_0,W_1}, and \eqref{Carnot}).
The open circles and triangles, respectively, express the work, absorbed heat, and efficiency obtained by Eq. \eqref{master} for $\epsilon =0.001$ and $\epsilon =0.01$, where even in the case of $\epsilon =0.01$ the deviation of the numerical result from the theoretical result in the adiabatic limit is invisible.
}
\label{fig:adiabaticWQeta}
\end{figure}

\subsection{Results in slow modulations}

\subsubsection{Results in the adiabatic limit}
\begin{figure}
\centering
\includegraphics[clip,width=16cm]{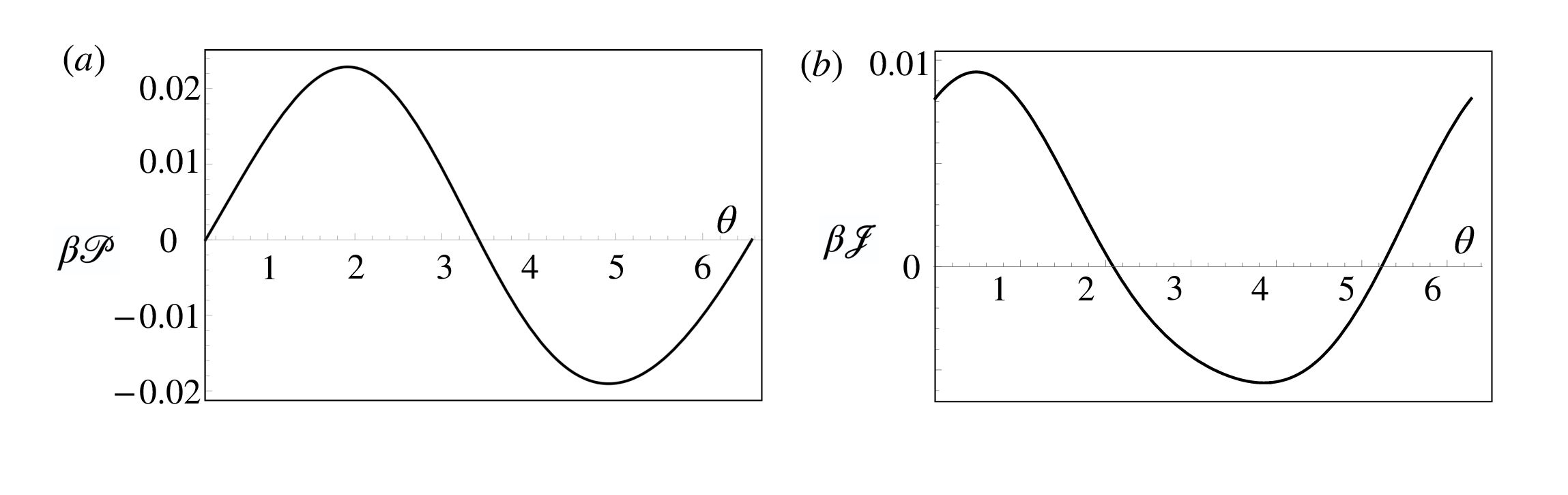}
\caption{
(a) Plot of $\mathscr{P}^{(0)}(\theta)$ (Eq.~\eqref{P_0,P_1}) versus $\theta$ for $\delta=0$.
(b) Plot of $\mathscr{J}^{(0)}(\theta)$ (Eq.~\eqref{J_geometric}) versus $\theta$ for $\delta=0$.
We commonly use the parameters $r=0.9$, $\beta U_0=0.1$, $\beta \epsilon_0=0.1$, $\theta_\mathrm{ini}=2\pi$, and $\beta \bar \mu =0.1$.   
}
\label{fig:PandQdot}
\end{figure}

For the explicit calculation for thermodynamic variables, we set $r:= r_\lambda=r_\mu=0.9$ and fix the parameters $\beta\epsilon _0 =\beta \overline{\mu}= \beta U_0 = 0.1$, where $\epsilon _0$ and $U_0$ are introduced in Eqs.\ \eqref{H_s} and \eqref{U(theta)}, respectively. 
Figure \ref{fig:adiabaticWQeta} plots the work \eqref{def:W}, the heat \eqref{def:Q}, and the efficiency \eqref{Carnot} in the adiabatic limit as the solid lines using the adiabatic $\hat{\rho}^\mathrm{ss}(\theta)$, where the integrals are evaluated numerically. 
Open circles and triangles in Fig.~\ref{fig:adiabaticWQeta}, respectively, denote the work, absorbed heat, and efficiency obtained by Eq. \eqref{master} for $\epsilon =0.001$ and $\epsilon =0.01$.
Remarkably, numerical solutions of Eq.~\eqref{master} for $\epsilon =0.001$ and $\epsilon =0.01$ are indistinguishable from the corresponding theoretical results in the adiabatic limit.
Moreover, these results can be reproduced using the analytic expressions Eqs.~\eqref{Wapprox} and \eqref{B4} in the high-temperature limit.
If we restrict our interest to the Anderson model in the adiabatic limit under the wide-band approximation as in this section, we can prove $W^{(0)}\le 0$ as shown in Appendix~\ref{app:proofs_efficiency}.

In Fig.~\ref{fig:PandQdot}, we plot $\theta$ dependence of theoretical $\mathscr{P}^{(0)}(\theta)$ (a) by Eq.~\eqref{P_0,P_1} and $\mathscr{J}^{(0)}(\theta)$ (b) by Eq.~\eqref{J_geometric} for $\delta=0$ in the adiabatic limit. 
As can be seen in Fig. \ref{fig:PandQdot}(b), the heat current $\beta\mathscr{J}^{(0)}(\theta)$ deviates from a sinusoidal curve, which generates non-zero heat.
Although $\mathscr{P}^{(0)}(\theta)$ appears to be a sinusoidal function of $\theta$, it also deviates slightly from the sinusoidal curve, generating non-zero work.

\subsubsection{Results in non-adiabatic regime}

\begin{figure}
\centering
\includegraphics[clip,width=18cm]{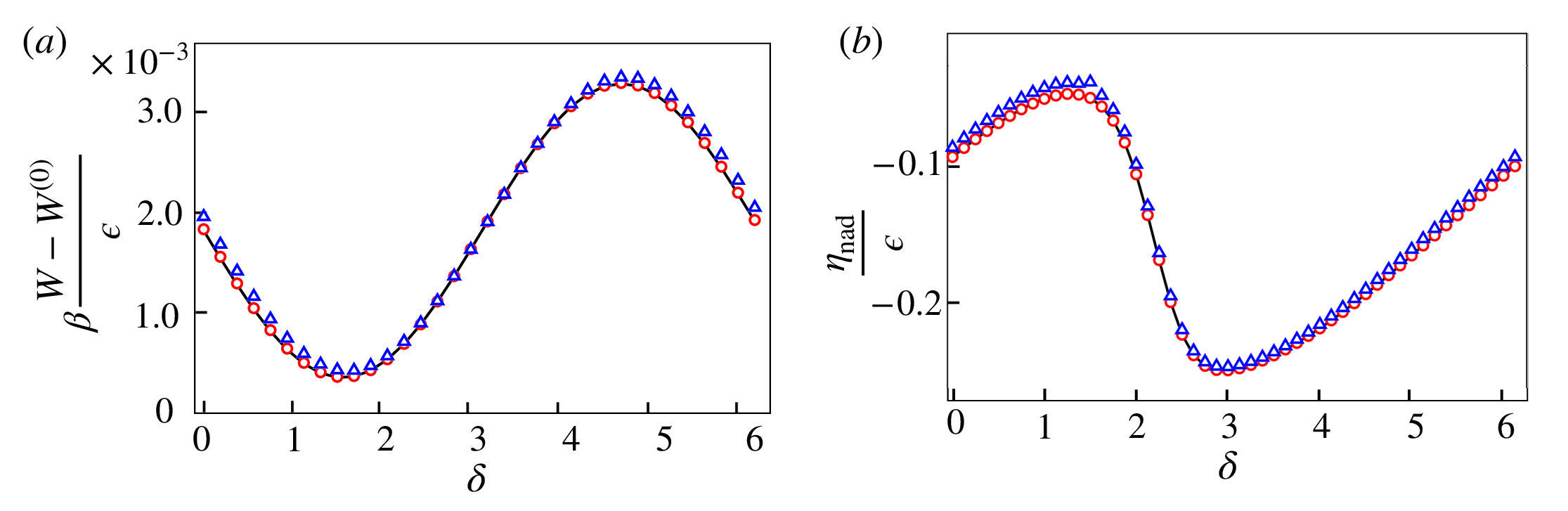}
\caption{
(a) Plots of $\beta ({W-W^{(0)}})/{\epsilon}$ versus $\delta$ for $r=0.9$, $\beta U_0=0.1$, $\beta \epsilon_0=0.1$, $\theta_\mathrm{ini}=2\pi$, and $\beta \bar \mu =0.1$.
The open red circles and blue triangles represent $(W-W^{(0)})/\epsilon$ obtained by the numerical solution of Eq.~\eqref{master} for $\epsilon=0.01$ and $\epsilon=0.1$, respectively.
The solid line represents $W^{(1)}$ via numerical integration of Eqs.~\eqref{W_0,W_1} using $|\hat{\rho}^{(1)}(\theta)\rangle$.
(b) Plots of $\eta_{\rm nad}/\epsilon$ versus $\delta$ for the same set of parameters.
The open red circles and blue triangles represent $\eta_{\rm nad}/\epsilon$ obtained by the numerical solution of Eq.~\eqref{master} for $\epsilon=0.01$ and $\epsilon=0.1$, respectively.
The solid line represents $\eta^{(1)}$ obtained via numerical integration of Eq. \eqref{eta^{(1)}} using $|\hat{\rho}^\mathrm{ss}(\theta)\rangle$ and $|\hat{\rho}^{(1)}(\theta)\rangle$.
}
\label{workdeltadep}
\end{figure}

In the non-adiabatic regime, we can use Eq.~\eqref{rho1_def_re} for the time evolution of the density matrix.
Using this, we can calculate the work, power, heat, and efficiency in our protocol to verify the validity of the theoretical expression for $\epsilon \ll 1$. 
Figure~\ref{workdeltadep} plots $W^{(1)}$ and $\eta^{(1)}$ against $\delta$. 
The solid lines represent $W^{(1)}$ and $\eta^{(1)}$, respectively, obtained by Eqs.~\eqref{W_0,W_1} and \eqref{eta^{(1)}}, and open circles and triangles, respectively, denote $(W-W^{(0)})/\epsilon$ and $\eta_\mathrm{nad}/\epsilon$ obtained by the numerical solution of Eq.~\eqref{master} for $\epsilon=0.01$ and $\epsilon=0.1$. 
The numerical results for $\epsilon=0.01$ perfectly agree with the theoretical results, while the numerical results for $\epsilon=0.1$ deviate a little from the theoretical results, in particular for small $\delta$ and large $\delta$.
This indicates that the contributions of $O(\epsilon^2)$ cannot be ignored for $\epsilon=0.1$.

\begin{figure}
\centering
\includegraphics[clip,width=18cm]{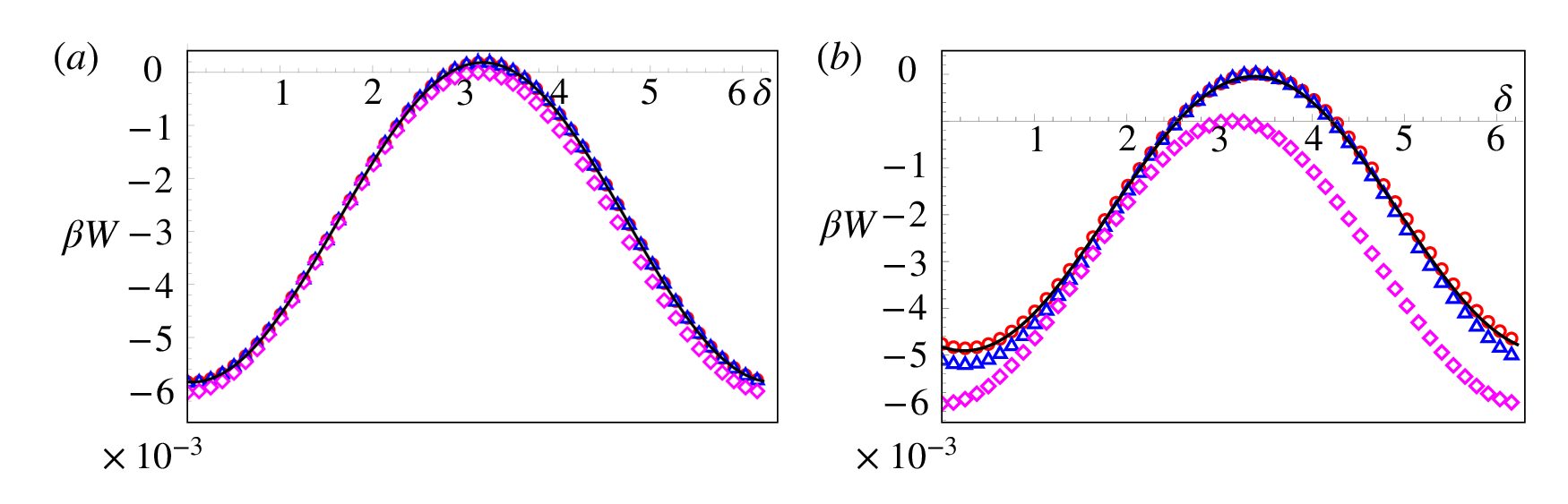}
\caption{
Plots of $\beta W$ for $\epsilon=0.1$ (a) and $\epsilon=0.5$ (b) with the fixing parameters $\beta \epsilon_0=0.1$, $\beta\bar \mu=0.1$, $\beta U_0=0.1$, and $r=0.9$. 
The solid line shows the result obtained by numerically solving the master equation Eq.~\eqref{master}. 
The open diamond, open triangle, and open circle are $W^{(0)}$,$W^{(0)}+\epsilon W^{(1)}$, and $ W^{(0)}+\epsilon W^{(1)}+\epsilon^2 W^{(2)}$, respectively.}
\label{workepsilonexpansion}
\end{figure}

We also plot $\delta$ dependence of the work for $\epsilon=0.1$ (Fig. \ref{workepsilonexpansion}~(a)) and $\epsilon=0.5$ (Fig.~\ref{workepsilonexpansion}~(b)) to validate the theoretical result in the $\epsilon$-expansion. 
The solid line shows the result obtained by the numerical solution of the master equation Eq.~\eqref{master}, while the open diamond, open triangle, and open circle 
are $W^{(0)}$,$W^{(0)}+\epsilon W^{(1)}$, and $ W^{(0)}+\epsilon W^{(1)}+\epsilon^2 W^{(2)}$, respectively, where $W^{(n)}:=\oint {\rm Tr}\left[\hat\rho^{(n)}\partial_\theta \hat H \right]$. 
We confirm that the expansion up to $O(\epsilon^3)$ shows excellent agreement even for $\epsilon=0.5$.

\begin{figure}
\centering
\includegraphics[clip,width=8cm]{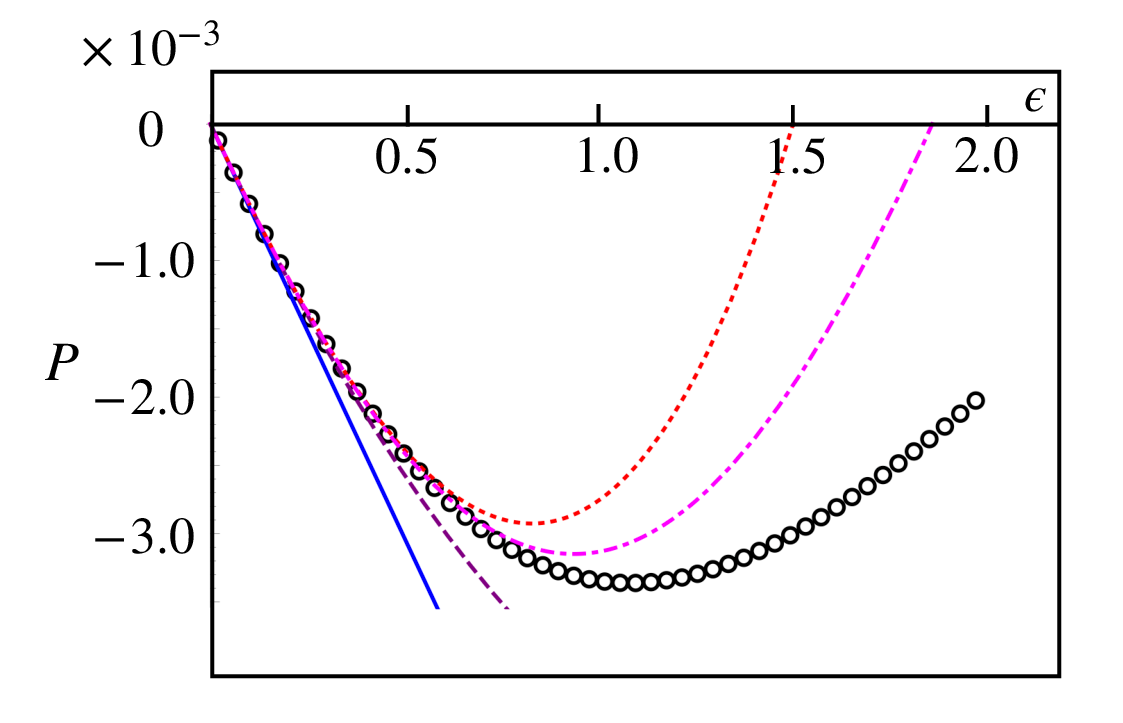}
\caption{
Plots of the power $P$ generated in a single cycle as functions of $\epsilon$ for $\delta=0$,  $\beta \epsilon_0=0.1$, $\beta\bar\mu=0.1$, $\beta U_0=0.1$, and $r=0.9$. 
The open circles represent the result obtained by the numerical solution of the master equation. 
The blue solid line, purple dashed line, red dotted line, and magenta dot-dashed line are $\epsilon W^{(0)}$,  $\epsilon W^{(0)}+\epsilon^2 W^{(1)}$, $\epsilon W^{(0)}+\epsilon^2 W^{(1)}+\epsilon^3 W^{(2)}$, and $\epsilon W^{(0)}+\epsilon^2 W^{(1)}+\epsilon^3 W^{(2)}+\epsilon^4 W^{(3)}$, respectively. }
\label{powerepsilondepdelta0}
\end{figure}

Figure \ref{powerepsilondepdelta0} plots the averaged power $P:=\epsilon W$ against $\epsilon$, where the open circles express the power obtained by the numerical solution of Eq.~\eqref{master}. 
From this figure, we recognize the limitation of the adiabatic theory (the blue solid line) and the non-adiabatic theory presented in this paper (the brown dashed line), where the numerical solution of Eq.~\eqref{master} deviates from both theoretical results. 
We also plot higher-order expansions for comparison: the red dotted line and magenta dot-dashed line correspond to $\epsilon W^{(0)}+\epsilon^2 W^{(1)}+\epsilon^3 W^{(2)}$ and $\epsilon W^{(0)}+\epsilon^2 W^{(1)}+\epsilon^3 W^{(2)}+\epsilon^4 W^{(3)}$, respectively. 
Although our theory up to $O(\epsilon^4)$ captures the behavior until \(\epsilon \lesssim 1\), the deviation between the theoretical result and the numerical result is non-negligible for large $\epsilon$.
This is because we adopt the small $\epsilon$ expansion, which breaks down in the large-$\epsilon(>1)$ region. 
However, we stress that even our non-adiabatic theory qualitatively captures the existence of a peak in the power at a certain $\epsilon$.
This is of practical importance for finite-time thermodynamics.

\begin{figure}
\centering
\includegraphics[clip,width=16cm]{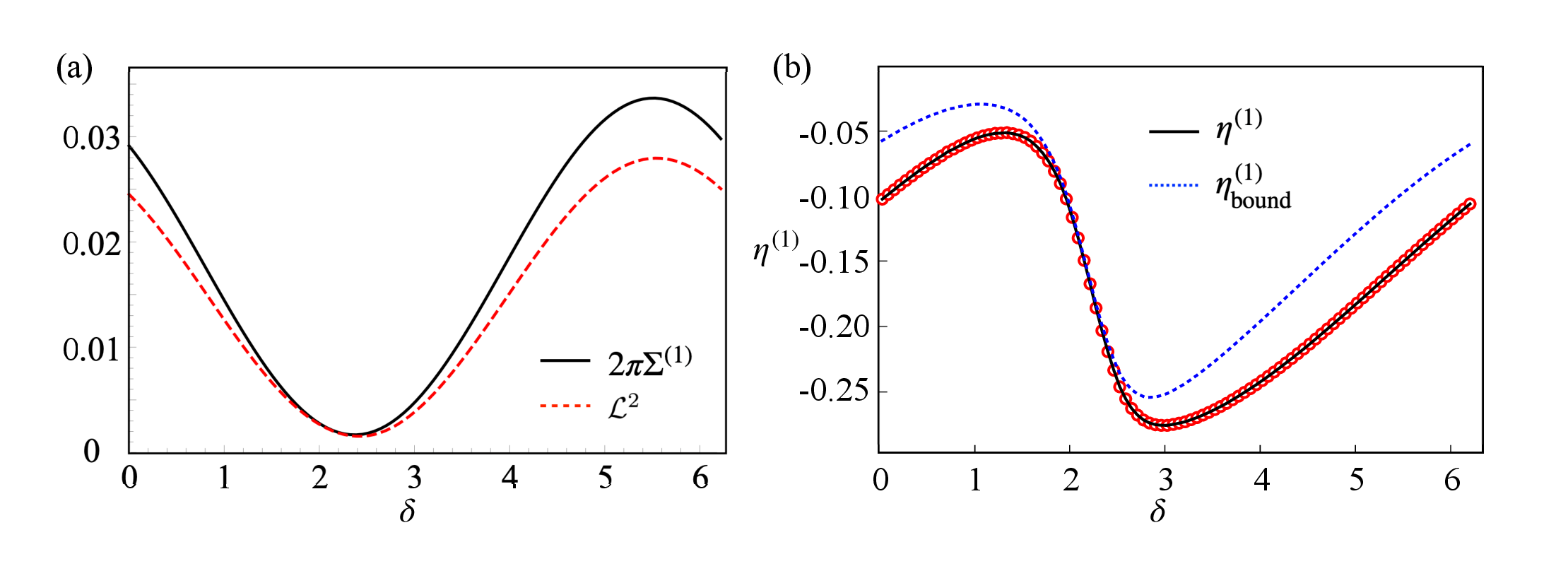}
\caption{
(a) Plots of the $\delta$-dependence of $2\pi\Sigma^{(1)}$ (the black solid line) and $\mathcal{L}^2$ (the red dashed line). 
(b) Plots of the $\delta$-dependence of $\eta^{(1)}$, where the black solid line, open circles, and blue dotted line represent the numerical result of $\eta_\mathrm{nad}/\epsilon$ in Eq.~\eqref{eta_nonad}, $\eta^{(1)}$, and the last expression in Eq.~\eqref{eta^{(1)}} for $\epsilon=0.001$. 
We commonly use $r=0.9$, $\beta U_0=0.1$, $\beta \epsilon_0=0.1$, and $\beta \bar \mu =0.1$. 
}
\label{fig:bounds}
\end{figure}

Figure~\ref{fig:bounds} validates the inequalities, Eqs.~\eqref{CS_re} and \eqref{eta_nonad}. 
In Fig.~\ref{fig:bounds} (a), we plot the $\delta$-dependence of $2\pi \Sigma^{(1)}$ and $\mathcal{L}^2$ for $r=0.9$, $\beta U_0=0.1$, $\beta \epsilon_0=0.1$, and $\beta \bar \mu =0.1$. 
We thus confirm the validity of Eq.~\eqref{CS_re}. 
Figure~\ref{fig:bounds} (b) shows the $\delta$-dependence of $\eta_\mathrm{nad}/\epsilon$ under the same set of parameters where the black solid line represents the numerical result of $\eta_\mathrm{nad}/\epsilon$ in Eq.~\eqref{eta_nonad} for $\epsilon=0.001$, the red open circles and the blue dotted line express $\eta^{(1)}$, and the last expression in Eq.~\eqref{eta^{(1)}}, respectively.
This result supports the inequality in Eq.~\eqref{eta^{(1)}}.
We also note that $\eta$ remains smaller than $\eta_{\rm ad}$ over the entire range for finite driving speeds.
This result is consistent with the theoretical proof in Appendix~\ref{app:proofs_efficiency}.

\subsection{The result in the fast modulations}

\begin{figure}
\centering
\includegraphics[clip,width=16cm]{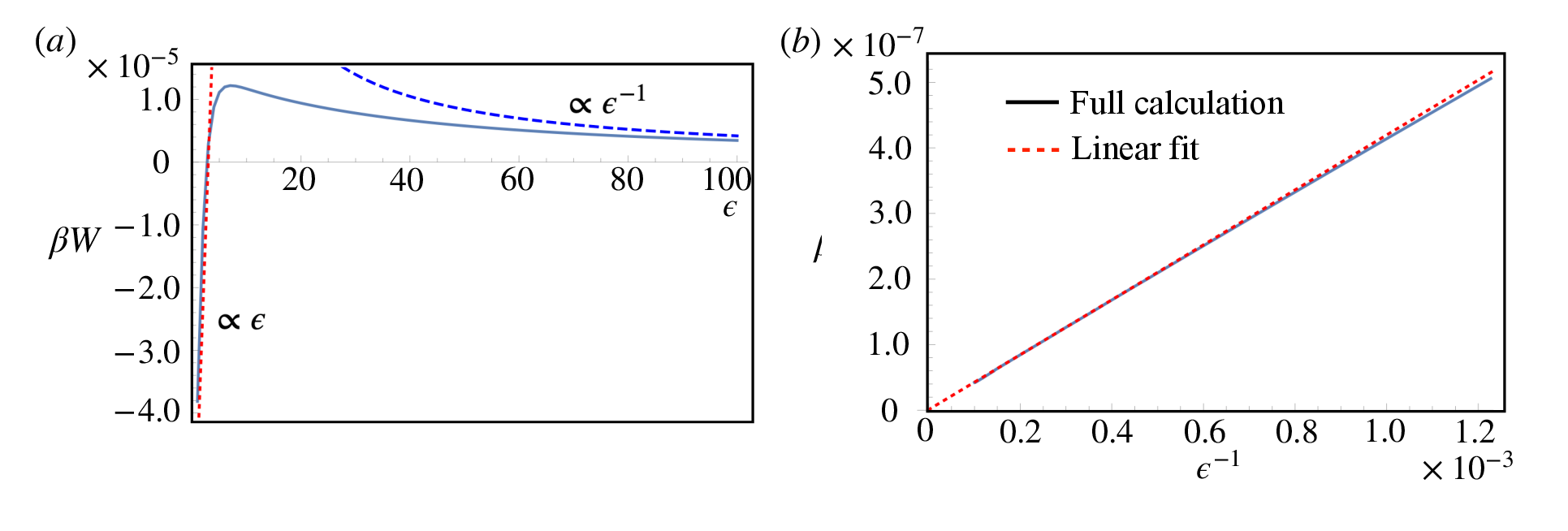}
\caption{(a) The work generated in a single cycle as a function of $\epsilon$ for $\delta=0$, $\beta\mu=0.1$, $\beta U_0=0.1$, and $r=0.9$. 
We also plot the linear fit in the vicinity of $\epsilon=0$ (dotted line) and the dashed line represents an $\epsilon^{-1}$ fit to the data. 
(b) Plot of $\beta W$ a as function of $\epsilon^{-1}$. 
}
\label{workepsilondepdelta0large}
\end{figure}

Figure~\ref{workepsilondepdelta0large} plots the work $W$ generated in a single cycle as a function of $\epsilon$ for $\delta=0$ with $\beta\epsilon_0=0.1$, $\beta\mu=0.1$, $\beta U_0=0.1$, $r=0.9$. 
As shown in Fig.~\ref{workepsilondepdelta0large}(a), the work takes a negative value in the adiabatic limit, increases linearly for small $\epsilon$, and reaches zero at a certain $\epsilon$.
Then, $W$ takes a maximum at another $\epsilon$, and it gradually decreases with $\epsilon$ (see Fig.~\ref{workepsilondepdelta0large}(a)), where the red dotted line and the blue dashed line are $\epsilon$ and $\epsilon^{-1}$, respectively.
The simulation confirm the validity of Eq.~\eqref{W_fastlimit} and the positivity of $W^{(-1)}$ in Eq.~\eqref{W-}, which is proven in Appendix \ref{app:proof_W_minus_one} (see Fig.~\ref{workepsilondepdelta0large}(b)).
Although we are interested in the efficiency $\eta$ as a thermodynamic engine for fast modulation cases, the work is always positive for large $\epsilon$.
Thus, we cannot extract the work for large $\epsilon$, and the efficiency $\eta$ is ill-defined.

\section{Discussion}\label{discussion}

In this paper, we have demonstrated that a periodically driven open quantum system can operate as an isothermal engine whose performance is governed by the geometry of the steady state rather than the conventional BSN curvature. While our concrete verification relies on the Anderson impurity model in the sequential tunneling regime, where off-diagonal density matrix elements decouple, a crucial next step is to address the role of many-body effects and electronic correlations more comprehensively. 
In the present wide-band approximation, the Coulomb interaction $U(\theta)$ acts primarily as an energetic barrier modulating the instantaneous steady-state populations. 
However, in deeper quantum regimes, such as at low temperatures where the Kondo effect dominates, strong electron-electron interactions give rise to emergent many-body resonances that dynamically hybridize with the leads. 
In such cases, the steady-state manifold itself inherits a highly non-trivial topological and geometric structure that cannot be described by a master equation. 
Extending our superoperator formulation to encompass these correlation-driven regimes will not only clarify how many-body states modify the thermodynamic metric tensor $g_{\mu\nu}(\Lambda)$ and efficiency corrections $\eta^{(1)}$, but also reveal whether strong interaction effects can be exploited to optimize work extraction and minimize finite-speed dissipation in nanoscale quantum devices.  

In conventional geometric pumping, the transported quantity per cycle is
determined by the BSN curvature, and the corresponding current is proportional to the operation speed.
In the strictly adiabatic limit, the instantaneous current vanishes,
while the integrated pumped quantity remains finite because it is given
by a geometric surface integral of the BSN curvature.

At first sight, the present engine might appear to belong to the same category as the Thouless pumping. 
However, the mechanism is fundamentally different.
The key point is that the work per cycle in our setup does not originate
from a geometric phase accumulated by the density matrix,
but from the steady-state structure of the Hamiltonian under cyclic
parameter modulation.
To see this explicitly, we note that the density matrix in the adiabatic regime can be written as Eq.~\eqref{rho_exp}.
The BSN curvature appears in the first-order correction
$|\hat{\rho}^{(1)}\rangle$, which indeed determines geometric pumping
currents proportional to $\epsilon$.
However, the work per cycle is given by Eq.~\eqref{W_0,W_1}, and its leading contribution arises from the instantaneous steady state
$|\hat{\rho}^{\mathrm{ss}}(\theta)\rangle$ itself.

Because the steady state depends on the electrochemical potentials and
Hamiltonian parameters, the cyclic modulation of these quantities
generates a finite work even in the limit $\epsilon \to 0$.
In this limit the BSN pumping current vanishes,
but the geometric integral of
$\mathrm{Tr}[\hat{\rho}^{\mathrm{ss}}(\theta)\partial_\theta \hat{H}(\theta)]$
remains finite.
Therefore, the extracted work is controlled by the parametric dependence
of the steady state rather than by the BSN curvature.
In other words, while the BSN curvature governs non-adiabatic pumping
currents, the present engine operates through a reversible modulation
of the steady-state manifold.
The geometric structure enters through the steady-state response
$\partial_\theta \hat{\rho}^{\mathrm{ss}}$,
not through the curvature of the BSN connection.
This distinction explains why the engine remains operative in the
strictly adiabatic limit and why its efficiency approaches its maximum value
as $\epsilon \to 0$.

Furthermore, it is instructive to position our findings within the broader context of nonlinear transport and thermoelectricity in quantum dot systems. 
The subtle roles of nonlinear responses and scattering mechanisms have been intensively investigated in literature; for instance, the magnetic field symmetries and the impact of elastic and inelastic scattering on nonlinear transport have been clarified based on an Aharonov-Bohm interferometer containing quantum dots \cite{Bedkihal2013}. 
In addition, the thermodynamic performance and efficiency optimization of nanoscale devices have been discussed in the context of quantum thermoelectrics under non-equilibrium electrons \cite{Volovichev2025}. 
While these studies focus on different driving mechanisms or transport regimes, integrating their insights—such as the effects of elastic/inelastic scattering and non-equilibrium electronic environments—into our superoperator formalism would provide a promising framework to further optimize the efficiency and power of the geometric isothermal engine.

\section{Concluding Remarks}\label{sec:conclusion}

In this paper, we have successfully extended the geometric thermodynamics formulated in Refs.~\cite{Brandner-Saito,engine,Hino2021} to a periodically driven system coupled to two reservoirs whose electrochemical potentials are slowly modulated. 
Our engine under periodic modulations corresponds to the Carnot cycle in standard thermodynamics in the adiabatic limit.
While our analysis is based on the formulation for quantum pumping, the explicit calculation in Sec.~\ref{sec:results} is still quasi-classical. 
We applied the formulation to the Anderson model within the wide-band approximation to obtain the explicit values of one-cycle averaged entropy production, thermodynamic length, adiabatic work, and effective efficiency after the system reaches a geometric (cyclic) state.
This is a natural extension of our previous work~\cite{Yoshii22}, which analyzed the initial relaxation to reach the geometric state.

So far, we have ignored the cost of manipulating the parameters as in standard thermodynamics. 
Of course, this assumption is not useful when considering the possibility of implementing a geometric thermodynamic engine. 
Although we do not know the correct cost, we evaluate the operation cost by assuming it equals the Joule heat.
For $\delta \neq 0$, the Joule heat can be estimated as the product of the dynamic current and the electrochemical potential difference. 
If $\delta=0$, we may use the Green-Kubo formula to evaluate the conductance, and thus we can estimate the cost.
Since our cost evaluation is far from perfect, we should compare our method with another approach, such as Ref.~\cite{Baruah2025}, as our future task.

Our future tasks are as follows:
(i) 
Although we considered the circular modulation of the parameters for simplicity in this study, the optimal path would be determined by the geodesic equation in the curved space characterized by the metric given by the Hessian matrix \cite{Abiuso}.
We also aim to determine the optimal parameter modulation path for our geometric thermodynamic engine using the geodesic equation. 
(ii) Although we have analyzed a quantum system, our treatment in Secs.~\ref{sec:appli} and \ref{sec:results} is still quasi-classical. 
In particular, the Anderson model used in this paper exhibits the Kondo effect at low temperatures as a quantum effect.
Unfortunately, the master equation approach has so far failed to describe the Kondo effect.
Therefore, this will be an important future task.
Furthermore, Viebahn et al.~demonstrated that interaction-driven topological pumping cannot be adiabatically connected to non-interacting limits~\cite{Viebahn}.
(iii) We also clarify the role of quantum coherence.
Brandner and Saito \cite{Brandner-Saito} showed that quantum coherence reduces the performance of slowly driven heat engines. 
On the other hand, it was shown that coherence can enhance the performance of heat engines in Refs.~\cite{coherence,Tajima2021}.
Therefore, we need to resolve the current confusing situation of whether quantum coherence leads to the enhancement of efficiency by using a fully quantum mechanical model. 
This may suggest the necessity of exploring quantum coherence effects in our model. 
On the other hand, Walter et al.~observed interaction-induced breakdown of Thouless pumping \cite{Walter}. 
This also suggests the importance of quantum coherence effects in adiabatic thermodynamic engines. 
(iv) We assumed that the master equation \eqref{master} is still Markovian even though parameter modulation is present.
However, this assumption may not be valid as indicated in Ref.~\cite{Mizuta21}. 
This is important because the modulation process might become non-CPTP if we use the Markovian dynamics described by Eq.~\eqref{master}.
Therefore, we should clarify the effect of non-Markovianity arising from the modulation. 
(v) Although we considered the reservoirs in our system with continuous spectra, we may extract topological features such as the Chern number if the conduction band is located near energy gaps in the reservoirs. 
Liu et al.~\cite{Liu} observed disorder-induced topological pumping, suggesting that similar effects could be explored in our model with gapped or disordered reservoirs. 
This would be an interesting direction for geometric thermodynamics.

\section*{Acknowledgements}

This work is partially supported by JSPS Grant-in-Aid for Scientific Research (KAKENHI Grant Nos.~JP25K07156 and JP26K06960). 
HH acknowledges the support of the Kyoto University Foundation and the JICA Friendship Program 2.0.
The authors acknowledge Ken Funo and Tan Van Vu for their critical reading of this manuscript and useful comments.
HH thanks Abhishek Dhar for the indication of Ref.~\cite{Wang2024}.

\newpage

\appendix

\section{Some detailed properties of general framework}\label{app:slow-driving}

In this appendix, we explain some general properties of the quantum master equation such as the outline of the perturbation method with a slowly modulated parameter in Appendix~\ref{slow_driving} and the mathematical description of the pseudo-inverse of the transition matrix in Appendix~\ref{app:pseudo-inverse}.

\subsection{Slow-driving perturbation}\label{slow_driving}

In this subsection, we explain the outline of the perturbation theory of the quantum master equation with a slowly modulated parameter $\epsilon$. 
First, we expand the solution of Eq. (\ref{master}) in terms of $\epsilon$ as
\begin{align}\label{p-expand}
    |\hat{\rho}(\theta)\rangle = \sum_{n=0}^{\infty} \epsilon^{n} |\hat{\rho}^{(n)}(\bm{\Lambda}(\theta))\rangle  
\end{align}
with $|\hat{\rho}^{(0)}\rangle=|\hat{\rho}^{\rm ss}\rangle$.
Since the normalization condition ${\rm Tr}[\hat{\rho}(\theta)] =1$ holds for any $\epsilon$, 
$|\hat{\rho}^{(n)}(\bm{\Lambda}(\theta))\rangle$ satisfies
\begin{align}
    &{\rm Tr}[\hat{\rho}^{\rm ss}(\bm{\Lambda}(\theta))] = 1, \\
    &{\rm Tr}[\hat{\rho}^{(n)}(\bm{\Lambda}(\theta))] =0
\end{align}
for $n\ge 1$.
Substituting them into Eq. (\ref{master}), we obtain Eq.~\eqref{steady_condition} and
\begin{align}
    \label{pn-eq}
    &\hat{K}(\bm{\Lambda}(\theta))|\hat{\rho}^{(n)}(\bm{\Lambda}(\theta))\rangle = \frac{d}{d\theta}|\hat{\rho}^{(n-1)}(\bm{\Lambda}(\theta))\rangle 
\end{align}
for $n\ge 1$,
By using the pseudo-inverse $\hat{K}^{+}(\bm{\Lambda}(\theta))$ of $\hat{K}(\bm{\Lambda}(\theta))$, Eq. (\ref{pn-eq}) can be written as
\begin{align}\label{pn}
    |\hat{\rho}^{(n)}(\bm{\Lambda}(\theta))\rangle 
    &= \hat{K}^{+}(\bm{\Lambda}(\theta)) \frac{d}{d\theta} |\hat{\rho}^{(n-1)}(\bm{\Lambda}(\theta))\rangle \notag \\
    &= \left(\hat{K}^{+}(\bm{\Lambda}(\theta)) \frac{d}{d\theta} \right)^{n} |\hat{\rho}^{\mathrm{ss}}(\bm{\Lambda}(\theta))\rangle.
\end{align}
Ignoring terms of $O(\epsilon^{2})$ and higher in Eq. (\ref{p-expand}), we obtain Eq. \eqref{rho1_def_re} of the main text by setting $n=1,2$.

\subsection{Pseudo-inverse of the transition matrix}\label{app:pseudo-inverse}

In this subsection, we introduce the pseudo-inverse $\hat{K}^{+}(\bm{\Lambda})$ of $\hat{K}(\bm{\Lambda})$, which satisfies following conditions 
\begin{align}
    \label{c1}
    &\hat{K}^{+}(\bm{\Lambda})\hat{K}(\bm{\Lambda}) =\hat{K}(\bm{\Lambda})\hat{K}^{+}(\bm{\Lambda}) = 1 - |\hat{\rho}^{\mathrm{ss}}(\bm{\Lambda})\rangle\langle 1|  , \\
    \label{c3}
    &\hat{K}^{+}(\bm{\Lambda})|\hat{\rho}^{\mathrm{ss}}(\bm{\Lambda})\rangle = 0, \\
    \label{c4}
    &\langle 1|\hat{K}^{+}(\bm{\Lambda}) =0.
\end{align}

Here, the eigen-equations are given by
\begin{align}\label{left_eigen_eq}
\langle \ell_{m}(\bm{\Lambda})| \hat{K}(\bm{\Lambda}) &=
\lambda_{m}(\bm{\Lambda}) \langle \ell_{m}(\bm{\Lambda})| \\
\hat{K}(\bm{\Lambda}) |r_{m}(\bm{\Lambda})\rangle &=
\lambda_{m}(\bm{\Lambda})  |r_{m}(\bm{\Lambda})\rangle , 
\label{right_eigen_eq}
\end{align} 
We note that $\lambda_{0}(\bm{\Lambda})=0$, then $|r_{0}(\bm{\Lambda})\rangle = |\hat{\rho}^{\mathrm{ss}}(\bm{\Lambda})\rangle$ and $\langle \ell_{0}(\bm{\Lambda})| = \langle 1|$.
For simplicity, we assume that these eigenstates do not degenerate throughout this paper.
Thus, we need to solve the eigenvalue problem \eqref{left_eigen_eq} or Eq.~\eqref{right_eigen_eq} to express the pseudo-inverse operator.

The definition of $\hat{K}^+$ in Eq.~\eqref{K+} satisfies the requirements of Eqs.\eqref{c1}-\eqref{c4}.
Indeed, using Eq.~\eqref{K+} we obtain
\begin{align}\label{proof_c1}
\hat{K}^+\hat{K}&=1
\sum_{m\ne0}\sum_n \frac{\lambda_n}{\lambda_m}
|r_m\rangle\langle \ell_m|r_n\rangle \langle \ell_n | 
\notag\\
&=\sum_{m\ne 0}\sum_{n\ne 0}\frac{\lambda_n}{\lambda_m}|r_m\rangle\langle\ell_n|\delta_{mn}
\notag\\
&=\sum_{n\ne 0} 
|r_n\rangle \langle \ell_n|+|r_0\rangle \langle \ell_0|-|\hat{\rho}^{\rm ss}\rangle \langle 1|  
\notag\\
&=1-|\hat{\rho}^{\rm ss}\rangle \langle 1| ,
\end{align}
where we have used $\lambda_0=0$ in the second line, $|r_0\rangle=|\hat{\rho}^{\rm ss}\rangle$ and $\langle \ell_0|=\langle 1|$ in the third line, 
and $\sum_n|r_n\rangle \langle \ell_n|=1$ in the last expression.
The second expression of Eq.~\eqref{c1} can be obtained by the parallel calculation to Eq.~\eqref{proof_c1}. 
Equation \eqref{c3} is the definition of the right zero eigenvector $|r_0\rangle=|\hat{\rho}^{\rm ss}\rangle$.
The proof of Eq. \eqref{c4} is also straightforward.
Indeed, substituting Eq.~\eqref{K+} into the left hand side of Eq.~\eqref{c4} we can write
\begin{align}
\langle 1|\hat{K}^+&=\sum_{m\ne 0}\frac{1}{\lambda_m}\langle 1|r_m\rangle \langle \ell_m|=0,
\end{align}
where we have used the orthogonal relation $\langle \ell_0|r_m\rangle:=\langle 1|r_m\rangle=0$ for $m\ne 0$.
Thus, Eq.~\eqref{K+} satisfies all requirements of the pseudo-inverse.

With the aid of Eq.~\eqref{c4} it is straightfoward to obtain
\begin{equation}\label{Tr[K+A]=0}
{\rm Tr}[\hat{K}^+\hat{A}]=\langle 1|\hat{K}^+\hat{A}|1\rangle=0
\end{equation}
for an arbitrary matrix $\hat{A}$.
Thus, we have the relation
\begin{equation}\label{Tr[D_mu]=0}
{\rm Tr}[\partial_\mu\partial_\nu \hat{\rho}^{\rm ss}]=
{\rm Tr}\left[\hat{K}^+\frac{\partial}{\partial \Lambda_\mu}
\left(\hat{K}^+\frac{\partial}{\partial \Lambda_\nu}\right) \hat{\rho}^{\rm ss}
\right]=0 .
\end{equation}
Multiplying $\hat{\rho}^{\rm ss}$ on the both side of the relation
\begin{equation}
\partial_\mu\partial_\nu\ln\hat{\rho}^{\rm ss}
=(\hat{\rho}^{\rm ss})^{-1}\partial_\mu\partial_\nu\hat{\rho}^{\rm ss}
-\partial_\mu \ln \hat{\rho}^{\rm ss}\cdot \partial_\nu \ln\hat{\rho}^{\rm ss}
\end{equation}
and take the trace with the aid of Eq.~\eqref{Tr[D_mu]=0} we obtain
\begin{equation}\label{Fisher=Hessian}
{\rm Tr}[\hat{\rho}^{\rm ss}\partial_\mu\partial_\nu \ln \hat{\rho}^{\rm ss}]
=-{\rm Tr}[\hat{\rho}^{\rm ss}\partial_\mu \ln \hat{\rho}^{\rm ss} \cdot \partial_\nu \ln \hat{\rho}^{\rm ss}] .
\end{equation}

\section{Proof of the Positivity of $W^{(-1)}$ in the Fast-Modulation Limit}
\label{app:proof_W_minus_one}

In this Appendix, we provide a formal proof for the non-negativity (and general positivity) of the leading-order work contribution $W^{(-1)}$ in the fast-modulation regime ($\epsilon \gg 1$), as given in Eq.~\eqref{W-} of the main text. 

The expression for the leading-order work correction over a closed parameter cycle is given by:
\begin{equation}
W^{(-1)} = \oint \langle \ell_0 | \dot{\hat{H}}(\theta) \left\{ \int_{\theta_{\text{ini}}}^{\theta} \hat{K}(\Lambda(\phi)) d\phi \right\} |\hat{\rho}^{\text{ss}}(\theta_{\text{ini}})\rangle,
\label{eq:W_minus_1_def}
\end{equation}
where $\theta$ is the cyclic phase variable, $\hat{H}(\theta)$ is the time-dependent Hamiltonian, $\hat{K}(\Lambda(\theta))$ is the instantaneous relaxation superoperator obeying local detailed balance, $\langle \ell_0 |$ represents the trace operation vector, and $|\hat{\rho}^{\text{ss}}(\theta_{\text{ini}})\rangle$ is the initial steady state frozen at the phase $\theta_{\text{ini}}$.

To analyze the sign of $W^{(-1)}$, we introduce the phase-dependent state deviation vector generated under the frozen-state approximation:
\begin{equation}\label{D2}
|\Delta \hat{\rho}(\theta)\rangle := \left\{ \int_{\theta_{\text{ini}}}^{\theta} \hat{K}(\Lambda(\phi)) d\phi \right\} |\hat{\rho}^{\text{ss}}(\theta_{\text{ini}})\rangle.
\end{equation}
Using this definition, Eq.~\eqref{eq:W_minus_1_def} can be compactly written as:
\begin{equation}
W^{(-1)} = \oint \langle \ell_0 | \dot{\hat{H}}(\theta) |\Delta \hat{\rho}(\theta)\rangle.
\end{equation}

We perform integration by parts with respect to the cyclic variable $\theta$ over the full loop $\theta \in [\theta_{\text{ini}}, \theta_{\text{ini}} + 2\pi]$. This yields:
\begin{equation}\label{D4}
W^{(-1)} = \Big[ \langle \ell_0 | \hat{H}(\theta) |\Delta \hat{\rho}(\theta)\rangle \Big]_{\theta_{\text{ini}}}^{\theta_{\text{ini}}+2\pi} - \oint \langle \ell_0 | \hat{H}(\theta) \frac{\partial}{\partial\theta} |\Delta \hat{\rho}(\theta)\rangle
= - \oint \langle \ell_0 | \hat{H}(\theta) \frac{\partial}{\partial\theta} |\Delta \hat{\rho}(\theta)\rangle
,
\end{equation}
where we have used the strict periodicity of driving parameters and the master equation over a complete cycle to obtain the last expression.
Namely, the boundary term vanishes identically ($\Delta \hat{\rho}(\theta_{\text{ini}}) = \Delta \hat{\rho}(\theta_{\text{ini}}+2\pi) = 0$). 
By applying the fundamental theorem of calculus to differentiate the inner integral, we obtain:
\begin{equation}
\frac{\partial}{\partial\theta} |\Delta \hat{\rho}(\theta)\rangle = \hat{K}(\Lambda(\theta)) |\hat{\rho}^{\text{ss}}(\theta_{\text{ini}})\rangle ,
\end{equation}
where we have used Eq.~\eqref{D2}.
Substituting this back into Eq.~\eqref{D4} simplifies the loop integral to a localized form:
\begin{equation}\label{D6}
W^{(-1)} = - \oint \langle \ell_0 | \hat{H}(\theta) \hat{K}(\Lambda(\theta)) |\hat{\rho}^{\text{ss}}(\theta_{\text{ini}})\rangle.
\end{equation}

From Eq.~\eqref{Q-} in the main text, the leading-order instantaneous heat current entering the system under this fast-modulation regime is defined as:
\begin{equation}
\mathscr{J}^{(-1)}(\theta) := \langle \ell_0 | \hat{H}(\theta) \hat{K}(\Lambda(\theta)) |\hat{\rho}^{\text{ss}}(\theta_{\text{ini}})\rangle.
\end{equation}
Thus, Eq.~\eqref{D6} directly expresses the cyclic relation between the leading-order fast work and heat:
\begin{equation}
W^{(-1)} = - \oint \mathscr{J}^{(-1)}(\theta) = - Q^{(-1)},
\end{equation}
which represents the fast-modulation limit of the first law of thermodynamics over a closed cycle.

To determine the sign of the integrand, we expand the frozen state $|\hat{\rho}^{\text{ss}}(\theta_{\text{ini}})\rangle$ around the moving, instantaneous steady state $|\hat{\rho}^{\text{ss}}(\theta)\rangle$ at the current phase $\theta$:
\begin{equation}
|\hat{\rho}^{\text{ss}}(\theta_{\text{ini}})\rangle = |\hat{\rho}^{\text{ss}}(\theta)\rangle + |\delta \hat{\rho}(\theta)\rangle,
\end{equation}
where $|\delta \hat{\rho}(\theta)\rangle$ represents the instantaneous lag from equilibrium. Since $|\hat{\rho}^{\text{ss}}(\theta)\rangle$ is the zero-right-eigenvector of the instantaneous master operator ($\hat{K}(\Lambda(\theta)) |\hat{\rho}^{\text{ss}}(\theta)\rangle = 0$), the heat current reduces to:
\begin{equation}
\mathscr{J}^{(-1)}(\theta) = \langle \ell_0 | \hat{H}(\theta) \hat{K}(\Lambda(\theta)) |\delta \hat{\rho}(\theta)\rangle.
\end{equation}

For any physically consistent Markovian generator governing sequential tunneling in open systems, $\hat{K}(\Lambda(\theta))$ satisfies the condition of local detailed balance. The spontaneous relaxation driven by $\hat{K}$ on any state deviation $|\delta \hat{\rho}(\theta)\rangle$ from the instantaneous equilibrium state must always be dissipative, meaning it reduces the expectation value of the energy deviation:
\begin{equation}\label{D11}
\langle \ell_0 | \hat{H}(\theta) \hat{K}(\Lambda(\theta)) |\delta \hat{\rho}(\theta)\rangle \le 0.
\end{equation}
As a result, the instantaneous heat current satisfies $\mathscr{J}^{(-1)}(\theta) \le 0$ at all points along the cycle where the system lags behind equilibrium. Integrating this inequality over the closed loop guarantees that the total heat entering the system is non-positive:
\begin{equation}
Q^{(-1)} = \oint \mathscr{J}^{(-1)}(\theta) \le 0.
\end{equation}

\begin{quote}
To prove Eq.~\eqref{D11}, we have added some extra explanations.
Substituting this back into the expression simplifies the loop integral to a localized form Eq.~\eqref{D6}.
To evaluate the sign of the integrand, we map the superoperator expression back to the conventional trace operation over the physical state space:
\begin{equation}
\langle \ell_0 | \hat{H}(\theta) \hat{K}(\Lambda(\theta)) |\hat{\rho}^{\text{ss}}(\theta_{\text{ini}})\rangle = \text{Tr}\left[ \hat{H}(\theta) \hat{K}(\Lambda(\theta)) \hat{\rho}^{\text{ss}}(\theta_{\text{ini}}) \right].
\end{equation}
We expand the frozen initial state $\hat{\rho}^{\text{ss}}(\theta_{\text{ini}})$ around the moving, instantaneous steady state $\hat{\rho}^{\text{ss}}(\theta)$ at the phase $\theta$ along the path:
\begin{equation}
\hat{\rho}^{\text{ss}}(\theta_{\text{ini}}) = \hat{\rho}^{\text{ss}}(\theta) + \delta \hat{\rho}(\theta),
\end{equation}
where $\delta \hat{\rho}(\theta) := \hat{\rho}^{\text{ss}}(\theta_{\text{ini}}) - \hat{\rho}^{\text{ss}}(\theta)$ is the instantaneous parameter lag. Since $\hat{\rho}^{\text{ss}}(\theta)$ is the right zero-eigenvector of the instantaneous master operator ($\hat{K}(\Lambda(\theta)) \hat{\rho}^{\text{ss}}(\theta) = 0$), the steady-state component vanishes identically under the trace. This isolates the deviation term:
\begin{equation}
\text{Tr}\left[ \hat{H}(\theta) \hat{K}(\Lambda(\theta)) \hat{\rho}^{\text{ss}}(\theta_{\text{ini}}) \right] = \text{Tr}\left[ \hat{H}(\theta) \hat{K}(\Lambda(\theta)) \delta \hat{\rho}(\theta) \right].
\end{equation}
For any Markovian generator $\hat{K}(\Lambda(\theta))$ obeying local detailed balance, the spontaneous relaxation acts dissipatively on any population imbalance $\delta \hat{\rho}(\theta)$, moving the system toward the instantaneous steady state. This relaxation inherently reduces the energy expectation value of the deviation, yielding the semi-definite inequality:
\begin{equation}
\text{Tr}\left[ \hat{H}(\theta) \hat{K}(\Lambda(\theta)) \delta \hat{\rho}(\theta) \right] \le 0.
\end{equation}

To establish the inequality in Eq.~\eqref{D11} explicitly, we project the trace into the instantaneous energy eigenbasis $\{|n\rangle\}$ of $\hat{H}(\theta)$ with eigenvalues $E_n(\theta)$. In this population representation, the trace reads:
\begin{equation}
\text{Tr}\left[ \hat{H}(\theta) \hat{K}(\Lambda(\theta)) \delta \hat{\rho}(\theta) \right] = \sum_{n} E_n(\theta) \sum_{m \neq n} \Big( W_{m \to n}(\theta) \delta \rho_m - W_{n \to m}(\theta) \delta \rho_n \Big),
\end{equation}
where $W_{m \to n}(\theta)$ are the transition rates and $\delta \rho_n = \langle n | \delta \hat{\rho}(\theta) | n \rangle$. Symmetrizing the double summation by swapping indices yields:
\begin{equation}
\text{Tr}\left[ \hat{H}(\theta) \hat{K}(\Lambda(\theta)) \delta \hat{\rho}(\theta) \right] = -\frac{1}{2} \sum_{n, m} \big( E_m(\theta) - E_n(\theta) \big) \Big( W_{m \to n}(\theta) \delta \rho_m - W_{n \to m}(\theta) \delta \rho_n \Big).
\end{equation}
Imposing local detailed balance with the isothermal reservoirs at temperature $T$, the energy gaps are strictly linked to the transition rates by $E_m(\theta) - E_n(\theta) = T \ln [W_{m \to n}(\theta)/W_{n \to m}(\theta)]$. Since the relaxation matrix transitions populations down the relative free-energy landscape toward the instantaneous steady state, every paired component in the summation satisfies the thermodynamic property $(x-y)\ln(x/y) \ge 0$. This ensures that the overall sum is non-negative, proving that the energy relaxation trace is semi-definite:
\begin{equation}
\text{Tr}\left[ \hat{H}(\theta) \hat{K}(\Lambda(\theta)) \delta \hat{\rho}(\theta) \right] \le 0.
\end{equation}

\end{quote}

Therefore, taking the negative of the cyclic heat, we arrive at the desired result:
\begin{equation}
W^{(-1)} \ge 0.
\end{equation}

This confirms that in the fast-modulation limit ($\epsilon \gg 1$), the external agent must perform positive work to overcome the continuous frictional dissipation caused by the system lagging behind the driving field, proving that work extraction as an engine is impossible in this regime.

\section{Eigenvalues and eigenvectors of the Anderson model}\label{app:eigen_modes}

In this appendix, we briefly summarize eigenvalues and eigenvectors of the Anderson model.
The Anderson model for a quantum dot should have the following four states (corresponding to $n=4$ in the previous section): doubly occupied, singly occupied with an up-spin, singly occupied with a down-spin, and empty.
Therefore, the density matrix should be expressed as a $4\times 4$ matrix. 
As is shown in Ref.~\cite{Yoshii13}, however, the density matrix of the quantum master equation of the Anderson model within the wide-band approximation is reduced to a four-component matrix $\hat{\rho}={\rm diag}[\rho_d,\rho_\uparrow,\rho_\downarrow,\rho_e]$, 
where $\rho_d,\rho_\uparrow,\rho_\downarrow$, and $\rho_e$ correspond to probabilities of the doubly occupied state, singly occupied state with up-spin, singly occupied state with down-spin, and empty state, respectively. 
This means that the model is not fully quantum-mechanical but quasi-classical.

NakaSince $\hat \rho$ is diagonal, $|\hat{\rho}\rangle$ also has only four components and the transition matrix $\hat{K}(\bm{\Lambda}(\theta))$ in Eq.~\eqref{master} in the wide band approximation is given by the $4\times 4$ matrix (see  Ref.~\cite{Nakajima15} for the derivation)
\begin{equation}\label{Ville_9_0806}
\hat{K}(\bm{\Lambda}(\theta))
=\begin{pmatrix}
-2f_-^{(1)} & f_+^{(1)} & f_+^{(1)} & 0 \\
f_-^{(1)}& -f_-^{(0)}-f_+^{(1)} & 0 & f_+^{(0)} \\
f_-^{(1)}& 0 & -f_-^{(0)}-f_+^{(1)}  & f_+^{(0)} \\
0 & f_-^{(0)} & f_-^{(0)} & -2f_+^{(0)} \\
\end{pmatrix} ,
\end{equation}
where we have introduced
\begin{align}
\label{f_+}
f_+^{(j)}:&=f_L^{(j)}(\mu^L,U)+f_R^{(j)}(\mu^R,U) \\
f_-^{(j)}:&=2-\{f_L^{(j)}(\mu^L,U)+f_R^{(j)}(\mu^R,U) \}
\label{f_-}
\end{align}
with the Fermi distribution
\begin{equation}\label{Fermi}
f_\alpha^{(j)}(\mu^\alpha(\theta),U(\theta)):=\frac{1}{1+e^{\beta (\epsilon_0+j U(\theta)-\mu^\alpha(\theta))}}
\end{equation}
in the lead $\alpha(=L~{\rm or}~ R)$.
Note that Eqs, \eqref{f_+} and \eqref{f_-} satisfy the relation 
\begin{equation}\label{sum_rule}
f_+^{(j)}+f_-^{(j)}=2
\end{equation}
for $j=0$ and 1.

It is straightforward to obtain the eigenvalues of $K(\bm{\Lambda}(\theta))$ in Eq.~\eqref{Ville_9_0806} as
\begin{align}
\lambda_0&=0, \label{0_eigen}\\
\lambda_1&=-(f_+^{(0)}+f_-^{(1)}), \label{1_eigen}\\
\lambda_2&=-(f_-^{(0)}+f_+^{(1)}), \label{2_eigen}
\\ 
\lambda_3&=-4.
\label{3_eigen}
\end{align}

The left and right eigenvectors corresponding to $\lambda_0=0$ in Eq.~\eqref{0_eigen} are given by
\begin{equation}\label{left_zero}
\langle \ell_0|=(1, 1, 1, 1) ,
\end{equation}
and 
\begin{equation}\label{right_zero}
|r_0\rangle 
=
\frac{1}{2(f_+^{(0)}+f_-^{(1)})}
\begin{pmatrix}
f_+^{(0)}f_+^{(1)} \\[0.5em]  
f_+^{(0)} f_-^{(1)}\\[0.5em]   
f_+^{(0)} f_-^{(1)}\\[0.5em]  
f_-^{(0)} f_-^{(1)}
\end{pmatrix},
\end{equation}
respectively.
Because of Eq.~\eqref{steady_condition} there is the trivial relation $|r_0\rangle =|\hat{\rho}^{\rm ss}\rangle$ for the diagonal element of the density matrix.
Note that $|r_0\rangle$ satisfies $\langle\ell_0|r_0\rangle={\rm Tr}\hat{\rho}^{\rm ss}=1$. 
The left and right eigenvectors corresponding to $\lambda_1$ in Eq.~\eqref{1_eigen} are given by
\begin{equation}
\label{left_1}
\langle \ell_1 |
=
2
\begin{pmatrix}
f_-^{(1)}, &
\frac{-f_+^{(0)}+f_-^{(1)}}{2}, &
\frac{-f_+^{(0)}+f_-^{(1)}}{2},&
-f_+^{(0)}
\end{pmatrix},
\end{equation}
and
\begin{equation}\label{right_1}
|r_1\rangle
=
\frac{1}{(f_+^{(0)}+f_-^{(1)})(f_-^{(0)}+f_+^{(1)})}
\begin{pmatrix}
f_+^{(1)}\\[0.5em]
\frac{-f_+^{(0)}+f_-^{(1)}}{2} \\[0.5em]
\frac{-f_+^{(0)}+f_-^{(1)}}{2}  \\[0.5em]
-f_-^{(0)}
\end{pmatrix}.
\end{equation}

The left and right eigenvectors corresponding to $\lambda_2$ in Eq.~\eqref{2_eigen} are
\begin{equation}\label{left_2}
\langle \ell_2 |=
2(0, 1, -1, 0),
\end{equation}
and
\begin{equation}\label{right_2}
|r_2\rangle
=\frac{1}{4}
\begin{pmatrix}
0\\
1\\
-1\\
0
\end{pmatrix}
,
\end{equation}
respectively.

The left and right eigenvectors corresponding to $\lambda_2$ in Eq.~\eqref{3_eigen} are 
\begin{equation}\label{left_3}
\langle \ell_3|
  =
\begin{pmatrix}
 f_-^{(0)}f_-^{(1)},&
-f_-^{(0)}f_+^{(1)},&
-f_-^{(0)}f_+^{(1)},&
 f_+^{(0)}f_+^{(1)}
\end{pmatrix},
\end{equation}
and
\begin{equation}\label{right_3}
|r_3\rangle
=\frac{1}{2(f_-^{(0)}+f_+^{(1)})}
\begin{pmatrix}
1 \\[0.5em]
-1\\[0.5em]
-1\\[0.5em]
1
\end{pmatrix}.
\end{equation}

\section{The work and heat in the high-temperature limit}\label{app:high_T}

In the main text, we consider thermodynamic behavior at a finite $\beta$.
Although the setup at finite $\beta$ is realistic, obtaining the analytic expressions is difficult.
However, if we focus on the high-temperature regime, $\beta U,\ \beta\mu, \beta\epsilon_0\ll 1$, we can obtain a closed expression for $W$ in the adiabatic regime $\epsilon\to 0$. 
For small $\beta$, one can adopt the expansion around $\beta=0$. 
For instance, the Fermi distribution function \eqref{Fermi} can be expanded in $\beta$ as follows. 
\begin{align}
f_\alpha^{(j)}&\simeq \frac{1}{2+\beta (\epsilon_0+j U(\theta)-\mu^\alpha(\theta))+\beta^2 (\epsilon_0+j U(\theta)-\mu^\alpha(\theta))^2/2}\notag \\
&\simeq \frac{1}{2}\left\{1+\frac{\beta (\epsilon_0+j U(\theta)-\mu^\alpha(\theta))}{2}+\left[\frac{\beta (\epsilon_0+j U(\theta)-\mu^\alpha(\theta))}{2}\right]^2\right\}^{-1}\notag\\
&\simeq \frac{1}{2}\left[1-\frac{\beta (\epsilon_0+j U(\theta)-\mu^\alpha(\theta))}{2}\right].\label{FermiApprox}
\end{align}

The adiabatic work for the Anderson model in the supervector formalism becomes 
\begin{align}
W^{(0)}&=\oint \,
\mathrm{Tr}\left[
\frac{\partial \hat{H}}{\partial \theta}\hat{\rho}^\mathrm{ss}(\theta)
\right]=-\oint U_0 r\sin\theta \,
\mathrm{Tr}\left[
\hat n_{\uparrow}\hat n_{\downarrow}\hat{\rho}^\mathrm{ss}(\theta)
\right]=-\oint U_0 r\sin\theta {\rho}^\mathrm{ss}_d(\theta),
\label{W0Anderson}
\end{align}
where 
$\hat{\rho}^\mathrm{ss}={\rm diag}[\rho_d^\mathrm{ss},\rho_\uparrow^\mathrm{ss},\rho_\downarrow^\mathrm{ss},\rho_e^\mathrm{ss}]$, 
where $\rho_d^\mathrm{ss},\rho_\uparrow^\mathrm{ss},\rho_\downarrow^\mathrm{ss}$, and $\rho_e^\mathrm{ss}$ correspond to probabilities of the doubly occupied state, singly occupied state with up-spin, singly occupied state with down-spin, and empty state for the steady state, respectively. 
Here $\rho_d^{\rm SS}$ is given by the first component of Eq.~\eqref{right_zero} and can be expanded by using Eqs.~\eqref{FermiApprox}, \eqref{f_+}, and \eqref{f_-}  as follows 
\begin{align}
\rho_d^{\rm SS}=&\frac{f_+^{(0)}f_+^{(1)}}{2(f_+^{(0)}+f_-^{(1)})}\notag\\
\simeq& \frac{1}{2}\frac{\left[1-\frac{\beta (\epsilon_0-\mu^L(\theta))}{4}-\frac{\beta (\epsilon_0-\mu^R(\theta))}{4}\right]\left[1-\frac{\beta (\epsilon_0+U-\mu^L(\theta))}{4}-\frac{\beta (\epsilon_0+U-\mu^R(\theta))}{4}\right]}{2-\frac{\beta (\epsilon_0-\mu^L(\theta))}{4}-\frac{\beta (\epsilon_0-\mu^R(\theta))}{4}+\frac{\beta (\epsilon_0+U-\mu^L(\theta))}{4}+\frac{\beta (\epsilon_0+U-\mu^R(\theta))}{4}} \notag \\
 \simeq& \frac{1}{4}
 \left[1-\frac{\beta (\epsilon_0-\mu^L(\theta))}{4}-\frac{\beta (\epsilon_0-\mu^R(\theta))}{4}\right]\left[1-\frac{\beta (\epsilon_0+U-\mu^L(\theta))}{4}-\frac{\beta (\epsilon_0+U-\mu^R(\theta))}{4}\right]\notag \\
 &\times \left[1+\frac{\beta (\epsilon_0-\mu^L(\theta))}{8}+\frac{\beta (\epsilon_0-\mu^R(\theta))}{8}-\frac{\beta (\epsilon_0+U-\mu^L(\theta))}{8}-\frac{\beta (\epsilon_0+U-\mu^R(\theta))}{8}\right]\notag\\ 
 \simeq &\frac{1}{4} -\frac{1}{32}\left[(\epsilon_0-\mu^L(\theta))+ (\epsilon_0-\mu^R(\theta))+3(\epsilon_0+U-\mu^L(\theta))+ 3(\epsilon_0+U-\mu^R(\theta))\right].
\end{align}

Thus, Eq.~\eqref{W0Anderson} yields 
\begin{align}
 W^{(0)}\simeq&  -\frac{\pi\bar \mu\beta r^2U_0 }{8}(1+\cos\delta). 
 \label{Wapprox}
\end{align}

\begin{figure}
\centering
\includegraphics[clip,width=16cm]{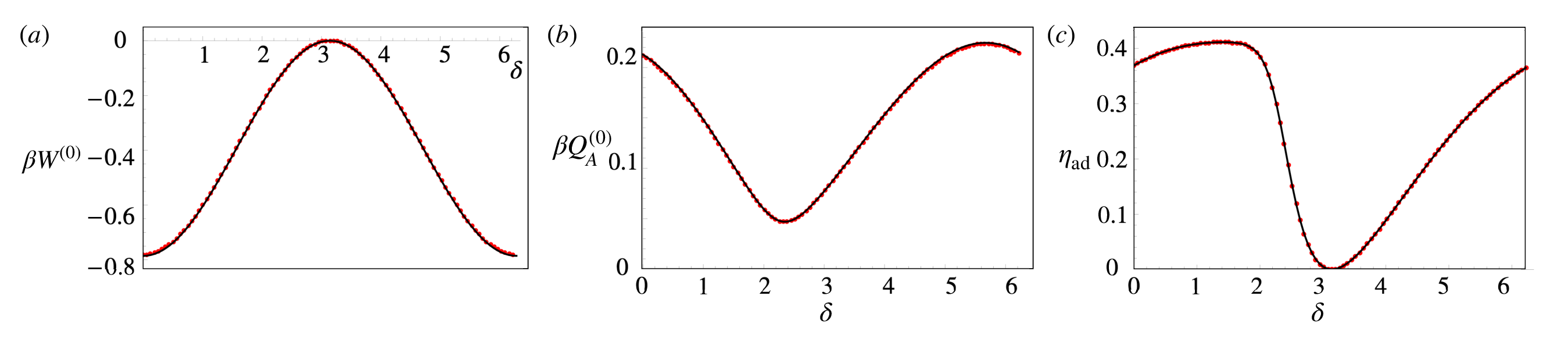}
\caption{
Plots of the adiabatic work (a), the adiabatic absorbed heat (b), and the adiabatic efficiency (c) as functions of $\delta$ for $r=0.9$, $\beta U_0=0.01$, $\beta \epsilon_0=0.01$, and $\beta \bar \mu =0.01$. 
The solid and dashed lines represent the exact solution and the result obtained by the lowest-order $\beta$-expansion, respectively. 
The plotted points represent the time evolution obtained via numerical integration. The solid lines show the results of the high-temperature expansion. For heat absorption, numerical integration was performed after applying the high-temperature expansion. For work, since the integration can be carried out analytically, the resulting analytical function is plotted. The efficiency is calculated as the ratio between the two.
}
\label{betadep}
\end{figure}

In Fig.~\ref{betadep}, we plot the $\delta$ dependence of the work for $\beta=1.0$, $0.1$, $0.01$, and $\beta=0.001$. 
The work becomes $0$ at $\delta=\pi$ in 0-th order in $\epsilon$ as seen from Eq.~\eqref{Wapprox}. 
This behavior appears since the work gain and the work lost in the left lead and those for the right lead are completely compensated when $\delta=0$ where $\mu_L=-\mu_R$ is held in any instance. 
The other interpretation can be drawn from the geometrical picture. 
In the case of $\delta=\pi$, the BSN curvature enclosed in the $\lambda-\mu_L$ plane and $\lambda-\mu_R$ has the same amplitude and opposite sign since $\mu_L=-\mu_R$. 
Thus, the BSN curvature enclosed by the trajectory becomes zero.

Similarly, the heat current can be approximated as 
\begin{align}\label{B4}
\mathscr{J}\simeq \frac{1}{16}\beta \{-2\epsilon_0U_0\sin x+(2\epsilon_0+U_0+U_0\cos x)(2\bar \mu[\cos x+\cos(x+\delta)]+3U_0\sin x)\}. 
\end{align}
The heat generated in a single cycle becomes 
\begin{align}
Q^{(0)}=\int_{\theta_{\rm in}}^{\theta_{\rm in}+2\pi}\mathscr{J}d\theta \simeq \frac{\pi\bar \mu\beta r^2U_0 }{8}(1+\cos\delta). 
 \label{Qapprox}
\end{align}
Eqs.~\eqref{Wapprox} and \eqref{Qapprox} satisfies the constraint $W^{(0)}+Q^{(0)}=0$ required for the adiabatic process. 
Calculating absorbed heat $Q^{(0)}_{\rm A}=\oint \mathscr{J}\Theta(\mathscr{J})d\theta$ cannot be explicitly done, and one needs to evaluate using numerics. 

\section{Proofs regarding the signs of adiabatic work and non-adiabatic efficiency corrections}
\label{app:proofs_efficiency}

In this appendix, we analyze the signs of the adiabatic work $W^{(0)}$ and the first-order non-adiabatic correction to the efficiency $\eta^{(1)}$. 
We first address why a general proof of $W^{(0)} \le 0$ is impossible for arbitrary open quantum systems, and subsequently provide a rigorous proof establishing $W^{(0)} \le 0$ and $\eta^{(1)} < 0$ for the Anderson model under the wide-band approximation.

\subsection{Sign of $W^{(0)}$ in a General Master Equation}
The adiabatic work per cycle is defined purely by the geometric trajectory across the instantaneous steady-state manifold:
\begin{equation}
W^{(0)} = \oint \text{Tr}\left[ \hat{\rho}^{\text{ss}}(\Lambda(\theta)) \frac{\partial \hat{H}(\Lambda(\theta))}{\partial \theta} \right] ,
\end{equation}
where $\hat{\rho}^{\text{ss}}(\Lambda)$ satisfies $\hat{K}(\Lambda)|\hat{\rho}^{\text{ss}}(\Lambda)\rangle = 0$. 

Mathematically, it is impossible to prove $W^{(0)} \le 0$ for an arbitrary transition matrix $\hat{K}(\Lambda)$ because the sign of $W^{(0)}$ depends entirely on the orientation and path of the parameter loop $\Lambda(\theta)$. Reversing the parametric orientation ($\theta \to -\theta$) flips the sign of $W^{(0)}$ without altering the underlying state manifold. Physically, a general master equation can describe devices acting as heat engines ($W^{(0)} < 0$), energy pumps ($W^{(0)} > 0$), or refrigerators. The Second Law bounds only the total cyclic entropy production $\Sigma^{(0)} = 0$ in this limit, imposing the constraint $W^{(0)} + Q^{(0)} = 0$ via the first law but providing no standalone constraint on the sign of $W^{(0)}$.

\subsection{Proof of $W^{(0)} \le 0$ for the Wide-Band Anderson Model}
We consider the Anderson impurity model in the wide-band limit with symmetric coupling ($V_L = V_R$). In this regime, the off-diagonal elements of the density matrix decouple from the diagonal populations, reducing the quantum master equation to a classical stochastic network over four charging states: empty ($|0\rangle$), spin-up ($|\!\uparrow\rangle$), spin-down ($|\!\downarrow\rangle$), and doubly occupied ($|d\rangle$).

Let the single-particle energy level $\epsilon_0$ be held constant, while the Coulomb repulsion $U(\theta)$ and the reservoir electrochemical potentials $\mu^L(\theta), \mu^R(\theta)$ are modulated cyclically according to the protocols:
\begin{align}
U(\theta) &= U_0 (1 + r_\lambda \cos\theta), \\
\mu^L(\theta) &= \bar{\mu}(1 + r_\mu \sin\theta), \\
\mu^R(\theta) &= \bar{\mu}(1 + r_\mu \sin(\theta + \delta)).
\end{align}
Since the time-dependence enters the system Hamiltonian solely through $U(\theta)\hat{n}_{\uparrow}\hat{n}_{\downarrow}$, the adiabatic work integral reduces to:
\begin{equation}
W^{(0)} = \oint \rho_d^{\text{ss}}(\theta) \frac{\partial U(\theta)}{\partial \theta} d\theta,
\end{equation}
where $\rho_d^{\text{ss}}(\theta):= \langle d | \hat{\rho}^{\text{ss}}(\theta) \rangle$ represents the instantaneous steady-state probability of double occupancy.

\begin{quote}
In a driven quantum dot, the adiabatic work per cycle is defined by the trace of the instantaneous steady state against the explicit time derivative of the system Hamiltonian Eq. \eqref{W_0,W_1}.
Let us look at the explicit structure of the Anderson impurity Hamiltonian on the dot in Eq.~\eqref{H_s}.
When we evaluate the parametric derivative $\partial_\theta \hat{H}(\theta)$, we must remember that the single-particle energy level $\epsilon_0$ is held perfectly constant throughout our protocol. 
Therefore, the only parameter that depends explicitly on $\theta$ is the Coulomb repulsion $U(\theta)$. 
Taking the derivative yields:
\begin{align}
    \frac{\partial \hat{H}(\theta)}{\partial \theta} = \frac{\partial U(\theta)}{\partial \theta} \hat{n}_\uparrow \hat{n}_\downarrow .
\end{align}
Now, let us evaluate the trace using the standard four charging states of the system ($|0\rangle, |\!\uparrow\rangle, |\!\downarrow\rangle, |d\rangle$):
For the empty state $|0\rangle$, the number operators yield $\hat{n}_\uparrow \hat{n}_\downarrow |0\rangle = 0$.
For the single-occupancy states $|\!\uparrow\rangle$ and $|\!\downarrow\rangle$, we have $\hat{n}_\uparrow \hat{n}_\downarrow |\!\sigma\rangle = 0$.
For the double-occupancy state $|d\rangle$, we have $\hat{n}_\uparrow \hat{n}_\downarrow |d\rangle = 1|d\rangle$.
When we expand the trace, the contributions from the empty and singly occupied states vanish identically because their eigenvalues under the operator $\hat{n}_\uparrow \hat{n}_\downarrow$ are zero:
\[\text{Tr}\left[ \hat{\rho}^{\text{ss}}(\theta) \frac{\partial \hat{H}(\theta)}{\partial \theta} \right] = \sum_{n \in \{0, \uparrow, \downarrow, d\}} \langle n | \hat{\rho}^{\text{ss}}(\theta) | n \rangle \langle n | \frac{\partial \hat{H}(\theta)}{\partial \theta} | n \rangle = \rho_d^{\text{ss}}(\theta) \frac{\partial U(\theta)}{\partial \theta} .\]
Thus, we only consider $\rho_d$.
\end{quote} 

Applying Stokes' theorem, we transform this line integral over the closed loop $\mathcal{C}$ into a surface integral over the parameter space area $\mathcal{S}$ enclosed by the cycle:
\begin{equation}
W^{(0)} = \iint_{\mathcal{S}} \left( \frac{\partial \rho_d^{\text{ss}}}{\partial \mu^L} d\mu^L \wedge dU + \frac{\partial \rho_d^{\text{ss}}}{\partial \mu^R} d\mu^R \wedge dU \right).
\end{equation}
In the sequential tunneling regime under the wide-band approximation, the steady-state occupancy $\rho_d^{\text{ss}}$ is structurally restricted by the Fermi-Dirac distributions of the leads. It obeys strict monotonic constraints across the entire operational parameter space:
\begin{enumerate}
    \item $\partial \rho_d^{\text{ss}}/\partial \mu^\alpha > 0 \quad (\alpha = L,R)$: Raising the chemical potential of either reservoir injects electrons into the dot, monotonically increasing the double-occupancy probability.
    \item $\partial \rho_d^{\text{ss}}/\partial U< 0$: Increasing the charging energy creates a Coulomb blockade penalty, monotonically suppressing double occupancy.
\end{enumerate}

Evaluating the explicit orientation of the wedge products under the prescribed driving protocols for a phase shift $\delta \ne \pi$ (where the chemical gates open as the interaction barrier decreases) yields:
\begin{equation}\label{E7}
W^{(0)} = -\pi r_\lambda r_\mu \bar{\mu} U_0 \left[ \left(\frac{\partial \rho_d^{\text{ss}}}{\partial \mu^L}\right) + \left(\frac{\partial \rho_d^{\text{ss}}}{\partial \mu^R}\right)\cos\delta \right].
\end{equation}
For any standard engine configuration for $\delta \ne \pi$, the term in brackets remains strictly positive due to the monotonic nature of the state manifold. 
Since $r_\lambda, r_\mu, \bar{\mu}, U_0 >0$, we arrive at the exact inequality:
\begin{equation}
W^{(0)} < 0 ,
\end{equation}
where we have used $\partial \rho_d^{\text{ss}}/\partial \mu^\alpha > 0 \quad (\alpha = L,R)$.

When the phase shift between the left and right chemical potentials reaches exactly $\delta = \pi$, the adiabatic work per cycle vanishes identically ($W^{(0)} = 0$). This behavior does not require the individual parameter derivatives of the steady-state density matrix to vanish; indeed, the physical response coefficients $\partial \rho_d^{\text{ss}}/\partial \mu^\alpha$ remain strictly positive across the entire parameter space. Instead, the vanishing of $W^{(0)}$ at $\delta = \pi$ is a direct consequence of the exact anti-phase modulation ($\cos\pi = -1$) under symmetric coupling ($V_L = V_R$), which enforces $\partial \rho_d^{\text{ss}}/\partial \mu^L = \partial \rho_d^{\text{ss}}/\partial \mu^R$ upon cyclic integration. 
See Eq.~\eqref{E7}.
As a result, the geometric contributions to the work gained from one lead and lost to the other completely compensate for each other over a full cycle, as explicitly verified via the low-order $\beta$-expansion in Appendix \ref{app:high_T}.

\subsection{Proof of $\eta^{(1)} < 0$ under standard engine conditions}
The first-order non-adiabatic correction to the efficiency $\eta^{(1)}$ is bounded by Eq.~(43):
\begin{equation}
\eta^{(1)} \le \frac{T\Sigma^{(1)} - W^{(1)} - T\mathcal{L}^2 / (2\pi)}{Q_A^{(0)}} + \frac{Q_A^{(1)} W^{(0)}}{(Q_A^{(0)})^2}.
\label{eq:eta1_bound}
\end{equation}
To evaluate the sign of $\eta^{(1)}$, we examine its exact identity derived from the expansion of $\eta = -W/Q_A$:
\begin{equation}
\eta^{(1)} = -\frac{1}{Q_A^{(0)}} \left[ W^{(1)} + \eta_{\text{ad}} Q_A^{(1)} \right],
\end{equation}
where $\eta_{\text{ad}} = -W^{(0)}/Q_A^{(0)}$. Using the cyclic relationship, Eq.~\eqref{Eq40}, $Q_\mathrm{rev}^{(1)} + W^{(1)} = T\Sigma^{(1)}$, where $Q^{(1)} = Q_A^{(1)} - Q_R^{(1)}$, and separating the irreversible dissipation component, the expression simplifies to:
\begin{equation}
\eta^{(1)} = -\frac{T\Sigma^{(1)}}{Q_A^{(0)}} - \frac{Q_A^{(1)}}{(Q_A^{(0)})^2} \left[ W^{(0)} + \mathcal{G} \right],
\end{equation}
where $\mathcal{G}$ represents the path-dependent boundary terms of the cyclic integrals. 
In a cyclic process, the total non-adiabatic heat exchanged with reservoir $\alpha$ is given by the integrated energy flux:
\begin{align}
Q_\alpha^{(1)} = \oint \text{Tr}\left[ \hat{H}(\theta) \hat{K}_\alpha(\theta) | \rho^{(1)}(\theta) \rangle \right] d\theta .    
\end{align}
By using the first-order master equation relationship $|\rho^{(1)}\rangle = \hat{K}^{-1} \frac{\partial}{\partial \theta} |\rho^{\text{ss}}\rangle$ and applying integration by parts over a full cycle, the explicit path-dependent boundary term $\mathcal{G}$ is isolated as:
\begin{align}
\mathcal{G} = \oint \text{Tr}\left[ \left( \hat{H}(\theta) - \sum_\alpha \mu^\alpha(\theta) \hat{N} \right) \frac{\partial \hat{\rho}^{\text{ss}}(\theta)}{\partial \theta} \right] d\theta ,
\end{align}
where $\hat{N} = \hat{n}_\uparrow + \hat{n}_\downarrow$ is the total number operator on the dot.

Note that $\mathcal{G}$ is negligible and vanishes identically for the wide-band Anderson model under the specific symmetric operations considered in our setup. 
There are two independent physical and structural reasons for this:
1) Symmetric Coupling and Single Quantum State: In our model, the coupling to the left and right reservoirs is symmetric ($V_L = V_R$), meaning that the chemical potentials $\mu^L(\theta)$ and $\mu^R(\theta)$ shift symmetrically around an average operating chemical potential $\bar{\mu}$. Because the total particle number $\hat{N}$ commutes with the charging states, the integration over a fully symmetric periodic loop ensures that the net accumulation of the $\mu^\alpha \hat{N}$ term integrates to zero over a closed cycle.
2) The reversible limit of the state manifold: The term $\mathcal{G}$ is structurally equivalent to a geometric boundary term representing the net difference between the instantaneous free energy paths. Because the system returns to its exact initial state at the end of the period ($\theta = 2\pi$), any total derivative or state-function boundary terms vanish upon cyclic integration:
\begin{equation}
    \oint \frac{\partial}{\partial \theta} \langle \hat{H} \rangle_{\text{ss}} = 0 .
\end{equation}
Therefore, in the derivation of the efficiency bound in Eq. \eqref{eta_nonad}, $\mathcal{G}$ can be safely set to zero. 

Under standard engine conditions, the system satisfies $W^{(0)} < 0$ (work is successfully extracted) and $Q_A^{(0)} > 0$ (finite heat is absorbed to run the engine). The first term on the right-hand side is strictly negative since the integrated entropy production is strictly positive ($\Sigma^{(1)} \ge \mathcal{L}^2/(2\pi) > 0$) due to finite-speed friction along the steady-state manifold. 

Because the wide-band Anderson model dictates strict monotonicity on the parameter derivatives ($\partial \rho_d^{\text{ss}}/\partial \mu^\alpha > 0$ and $\partial \rho_d^{\text{ss}}/\partial U < 0$), the geometric path configuration bounds the non-adiabatic heat fluctuations such that the dissipation term $-T\Sigma^{(1)} / Q_A^{(0)}$ dominates over the modulation correlation terms. Consequently, the efficiency degradation coefficient is forced to be strictly negative:
\begin{equation}
\eta^{(1)} < 0.
\end{equation}
This confirms that operating the wide-band Anderson model at a finite speed ($\epsilon > 0$) inevitably degrades the performance below the adiabatic limit.


\end{document}